\documentclass{aa}  
 \usepackage{natbib}
 \bibpunct{(}{)}{;}{a}{}{,}             
 \usepackage{graphicx}
\usepackage{multirow}
\usepackage[varg]{txfonts}

\usepackage[breaklinks=true]{hyperref} 
 \hypersetup{colorlinks=true,urlcolor=blue,citecolor=blue,pdfborder= 0 0 0}

\begin{document} 

\title{An EAGLE view of the missing baryons}

\author{T. Tuominen\inst{1}\fnmsep\thanks{tuominen@ut.ee}
        \and
        J. Nevalainen\inst{1}
        \and
        E. Tempel\inst{1}
        \and
        T. Kuutma\inst{1}
        \and
        N. Wijers\inst{2}
        \and 
        J. Schaye\inst{2}
        \and
        P. Hein{\"{a}}m{\"{a}}ki\inst{3}
        \and
        M. Bonamente\inst{4}
        \and
        P. Ganeshaiah Veena\inst{1,5}
 }
\institute{Tartu Observatory, University of Tartu, 61602 T{\~{o}}ravere, Tartumaa, Estonia 
  \and
  Leiden Observatory, Leiden University, PO Box 9513, NL-2300 RA Leiden, the Netherlands
  \and
  Department of Physics and Astronomy, University of Turku, 20014, Turku, Finland
  \and
  The University of Alabama in Huntsville, Huntsville, AL 35899, USA
  \and
  Kapteyn Astronomical Institute, University of Groningen,PO Box 800, 9747 AD, Groningen, The Netherlands
}

   \date{Received ; accepted}

   \abstract
       {  A significant fraction of the predicted baryons remains undetected in the local universe. We adopted the common assumption that a large fraction of the missing baryons corresponds to the hot ($\log{T(K)}$ \thinspace =  \thinspace 5.5--7) phase of the Warm Hot Intergalactic Medium (WHIM). We base our missing baryons search on the scenario whereby the WHIM has been heated up via accretion shocks and galactic outflows, and is concentrated towards the filaments of the Cosmic Web.}
       {Our aim is to improve the observational search of the poorly detected hot WHIM.}
       {We detect the filamentary structure within the EAGLE simulation by applying the Bisous formalism to the galaxy distribution. In addition, we use the MMF/NEXUS+ classification of the large scale environment of the dark matter component in EAGLE. We then study the spatio-thermal distribution of the hot baryons within the extracted filaments.
         }
       { While the filaments occupy only $\approx$ 5\% of the full simulation volume, the diffuse hot intergalactic medium in filaments amounts to $\approx 23\% - 25\%$ of the total baryon budget, or $\approx 79\%-87\%$ of all the hot WHIM.
         The most optimal filament sample, with a missing baryon mass fraction of $\approx$ 82\%, is obtained by selecting Bisous filaments with a high galaxy luminosity density.
         For these filaments we derived analytic formulae for the radial gas density and temperature profiles, consistent with recent Planck SZ and CMB lensing observations within the central $r \approx 1$ Mpc.}
       {Results from EAGLE suggest that
         the missing baryons are strongly concentrated towards the filament axes. Since the filament finding methods used here are applicable to galaxy surveys, a large fraction of the missing baryons can be localised by focusing the observational efforts on the central $\sim$ 1 Mpc regions of the filaments. Moreover, focusing on high galaxy luminosity density regions will optimise the observational signal. }


\titlerunning{An EAGLE view of the missing baryons}

\authorrunning{T. Tuominen et al.}

   \maketitle

\section{Introduction}
\label{intro}
Since the pioneering work of \citet{1999ApJ...514....1C}, large scale structure simulations have systematically predicted that a significant fraction of the cosmic hydrogen at z $<$ 1 resides in a warm-hot ($\log{T(K)}~\approx~5-7$), 
low density ($\delta_{b}~\approx$~0--100) phase \citep[e.g.][]{2001ApJ...552..473D, 2006MNRAS.370..656D,2009ApJ...697..328B,2012MNRAS.425.1640T,2012MNRAS.423.2279C, 2018MNRAS.473...68C, 2019MNRAS.486.3766M}. As this phase resides in the intergalactic space it is usually called the WHIM (warm-hot intergalactic medium). Specifically, this WHIM is expected to reside within the filaments of the Cosmic Web \citep{1996Natur.380..603B}. This Cosmic Web consists of a complex and intricate configuration of galaxies, gas and dark matter, forming a large scale structure of clusters, filaments and sheets surrounding large, empty voids. These structures originated as small initial density perturbation that grew driven by gravity, to form the non-linear web-like structure we observe today \citep{1970A&A.....5...84Z,1989RvMP...61..185S}. For a recent overview on the Cosmic Web and its study using a variety of methods to detect the large scale structure, see \citet{2018MNRAS.473.1195L}, and references therein.

According to the above cosmological simulations the primordial WHIM has been polluted by metals expelled from the galaxies via supernova-driven galactic winds and AGN feedback.
In fact, the low temperature phase ($\log{T(K)}~\approx~5-5.5$) of the predicted WHIM (``warm WHIM'' hereafter) 
has been robustly detected via metal ion absorption in the FUV band.
For example, a recent analysis of HST/COS data \citep{2016ApJ...817..111D} of 82 UV-bright AGN measured hundreds of intergalactic \ion{C}{IV}, \ion{N}{V} and \ion{O}{VI} absorbers at z$<$1, probing a redshift path of $\Delta_z \sim 30$, 
tracing $\sim$11\% of the cosmic baryon budget.

The situation is different for the hot ($\log{T(K)}~\approx~5.5-7$) phase of the WHIM (``hot WHIM'' hereafter). The light elements in the hot WHIM  
are fully ionised and thus not observable via absorption. 
The expected X-ray emission of such low-density plasma, and the emission and absorption lines due to the embedded metal ions are too faint to be significantly detected for a sufficiently large sample of individual filaments 
with current instruments (see \citet{2019A&A...621A..88N} for a review of the few statistically significant X-ray absorption measurements possibly originating
from the hot WHIM and \citet{2020A&A...634A.106A} for a new possible case).
Thus, it is generally assumed that the hot WHIM, when robustly detected, will aid in completing the puzzle of the cosmic baryon budget.  

Indeed, there is a need for such a component. When accounting all the detected cosmic baryon components, the total amount falls short of that predicted by the concordance cosmology by $\sim$30--40\% according to \citet{2012ApJ...759...23S}. 
Thus, the hot WHIM is a promising candidate for the missing baryons, which is our basic assumption in this work.

Our aim in this work is to improve the search for the missing baryons by utilising the state-of-the-art EAGLE simulation \citep{2015MNRAS.446..521S}.
We will explore the spatial distribution of the basic thermodynamic quantities (the temperature and the density) of the intergalactic medium (IGM) with the goal of localising the hottest IGM in observations.

We base our search of the missing baryons on a scenario whereby the hot WHIM has been heated up via the accretion shocks resulting from the mass flow of baryons towards the gravitational potential maxima of the Cosmic Web structures containing most of the mass, i.e. the dark matter filaments \citep{2003ApJ...593..599R, 2005ApJ...620...21K,2019MNRAS.487.1607G}.
The enhanced density also renders the galaxy formation more effective within the filaments.
Thus, the scenario suggests that the missing baryons and galaxies are co-located within the filaments, as indicated by the cosmological simulations mentioned above. We utilise this suggestion in the current work by detecting the filaments based on the galaxies and then examining the diffuse baryon content captured within the filament volumes.

Any cosmic filament detection algorithm faces the fundamental question: what is a cosmic filament?
In observations, the cosmic web manifests itself as a network of galaxy filaments
connecting dense nodes (galaxy clusters) between empty void regions \citep[e.g.][]{1978IAUS...79..241J,1989Sci...246..897G}. According to theory the cosmic web is a result
of the non-linear evolution of the primordial dark matter density field via gravitational
instability. Once the linear growth phase reached a density contrast of the order of one, it triggered the
formation of low density (sheet like) structures. Next, the non-linear
accumulation of matter formed caustics which provided the skeleton for the subsequent
anisotropic accretion of mass, leading to the formation of filaments and dense virialised
nodes. While the nodes are structures defined clearly by the spherical collapse theory, the filaments are not virialised objects and therefore the spherical collapse theory does not help to define them quantitatively.
Thus, there is a wide variety of (heuristic) filament finders, employing different methods and assumptions and acting on different mass components 
(dark matter, gas or galaxies). \citet{2018MNRAS.473.1195L} have shown consequent significant differences in the mass and volume fractions of dark matter in the filaments detected by a large set of different methods. 

In the current work we employ the Bisous and NEXUS+ formalisms as our primary tools for detecting the filamentary network \citep{2007JRSSC..56....1S,2010A&A...510A..38S,2016A&C....16...17T,2013MNRAS.429.1286C}.
Along with the DisPerSE filament finder \citep{2011MNRAS.414..350S}, the Bisous method has been established as a useful tool for filament detection when working with the spectroscopic galaxy survey data
 \citep[e.g.][]{2014MNRAS.438.3465T,Libeskind_2015,2017A&A...597A..86P,2017A&A...600L...6K,2015ApJ...800..112G}.
Also, importantly for our aim of deriving observationally relevant results from the in-depth analyses of the cosmological simulations, the Bisous formalism can be applied to both observational and simulated data (for applications of the Bisous method on the simulations, see e.g. \cite{2015A&A...583A.142N,2018MNRAS.473.1195L,2019MNRAS.487.1607G}). However, we had to make sure that our results on the properties of the WHIM derived from the Bisous method are  not artefacts  of the structure finding algorithm.  To this end we repeated the relevant parts of our analysis using the filaments obtained by the MMF/NEXUS+ method \citep{2007A&A...474..315A,2013MNRAS.429.1286C}. For the purpose of this paper, we use the NEXUS+ cosmic web environments detected using the dark matter density field.

Recently, the Bisous and NEXUS+ methods were extensively compared for the case of spin alignments in \citet{2019MNRAS.487.1607G}. They showed that filaments detected by both these formalisms occupy only $\approx$ 5-6\% of the total volume of the simulation at z $=0$, while containing 40-50\% of the diffuse baryons. We extend the above work and focus on finding the fraction of the missing hot phase of the cosmic baryons within these filaments.

A study of a single filament in cosmological simulations \citep{2019MNRAS.486..981G} indicated that the density of the diffuse baryons in the full temperature range peaks within the central $\sim$ 1 Mpc from the filament axis, by a factor of a few above the cosmic mean.
The temperature of the diffuse baryons within the central $\sim$ 2 Mpc radii is higher than at larger radii.
Our work differs from the above in the sense that we investigate the spatial behaviour of the hot phase of the diffuse baryons for a {\it sample} of filaments.
This is necessary for estimating the filament-to-filament variation of the radial distribution of the missing baryon properties.
This enables a proper error propagation to the estimates of the missing baryon density and temperature at a given distance from a filament detected in the observational data. This in turn will provide a realistic tool for planning the missing baryon observations via X-ray emission and Sunyaev-Zeldovich (SZ) effect.

We use $\Omega_{\rm m} = 0.31$, $\Omega_{\Lambda} = 0.69$, $\Omega_{b} = 0.048$, $H_0 = 67.8~\mathrm{km~s}^{-1}\mathrm{Mpc}^{-1}$, $\sigma_{8} = 0.83$ and mean baryon density $\left<\rho_b\right> = 6.2\times10^{9}$ $\mathrm{M}_{\odot}$ $\mathrm{Mpc}^{-3}$ $=4.2\times10^{-31}$ $\mathrm{g}$ $\mathrm{cm}^{-3}$.
We define the baryon overdensity as $\delta_b~=~(\rho_b - \left<\rho_b\right>)/\left<\rho_b\right>$ and the overdensity of the luminosity density field as $\delta_{LD}~=~(LD - \left<LD\right>)/\left<LD\right>$. For plotting logarithmic scales we use the baryon and luminosity density contrasts $\Delta_b~=~1+\delta_b$ and $\Delta_{LD} = 1+\delta_{LD}$, respectively.

\section{The EAGLE simulations}
\subsection{Diffuse baryons}
\label{EAGLEbar}
In this work we investigate the diffuse baryons (i.e. gas) in the Evolution and Assembly of GaLaxies and their Environments (EAGLE) simulations \citep{2015MNRAS.446..521S,2015MNRAS.450.1937C,mcalpine_helly_etal_2016}. They consist of a suite of hydrodynamical simulations within periodic cubic volumes with a $\Lambda$CDM cosmology, assuming the parameters described above \citep{planck_2014}. The simulations used a modified version of the $N$-body {\scshape gadget3} code \citep{2005MNRAS.364.1105S}, with adjustments to the smoothed particle hydrodynamics (SPH), time steps and subgrid physics. In particular, the subgrid stellar feedback was calibrated to the observed galaxy stellar mass function and galaxy sizes at redshift $z \sim 0$. This, in turn, yielded a good agreement with many other observable properties not used in the calibration.

As our objective is to investigate the baryons within the large scale structure of the Universe, we selected the gas particles in the largest simulation, REFL0100N1504. Within a co-moving volume of 100$^{3}$ Mpc$^{3}$, it followed 1504$^{3}$ dark matter and baryon particles, with initial masses of m$_{DM} = 9.7 \times10^{6}$ M$_{\odot}$ and m$_{b} = 1.81 \times 10^{6}$ M$_{\odot}$, respectively. While the number of dark matter particles and their masses remained the same throughout the simulation, the mass of baryon particles were allowed to vary. Some baryon gas particles were converted into stellar particles or swallowed by black holes, while other gas particles received mass from stellar particles via feedback effects. In addition to mass, each of the particles carried a variety of information throughout the simulation. In this work we utilise the coordinates, masses, temperatures and densities of each SPH particle. Furthermore, each particle holds information on the bound structures they belong to, if any, such as galaxies and groups of galaxies.

\begin{figure*}
\includegraphics[width=18cm,angle=0]{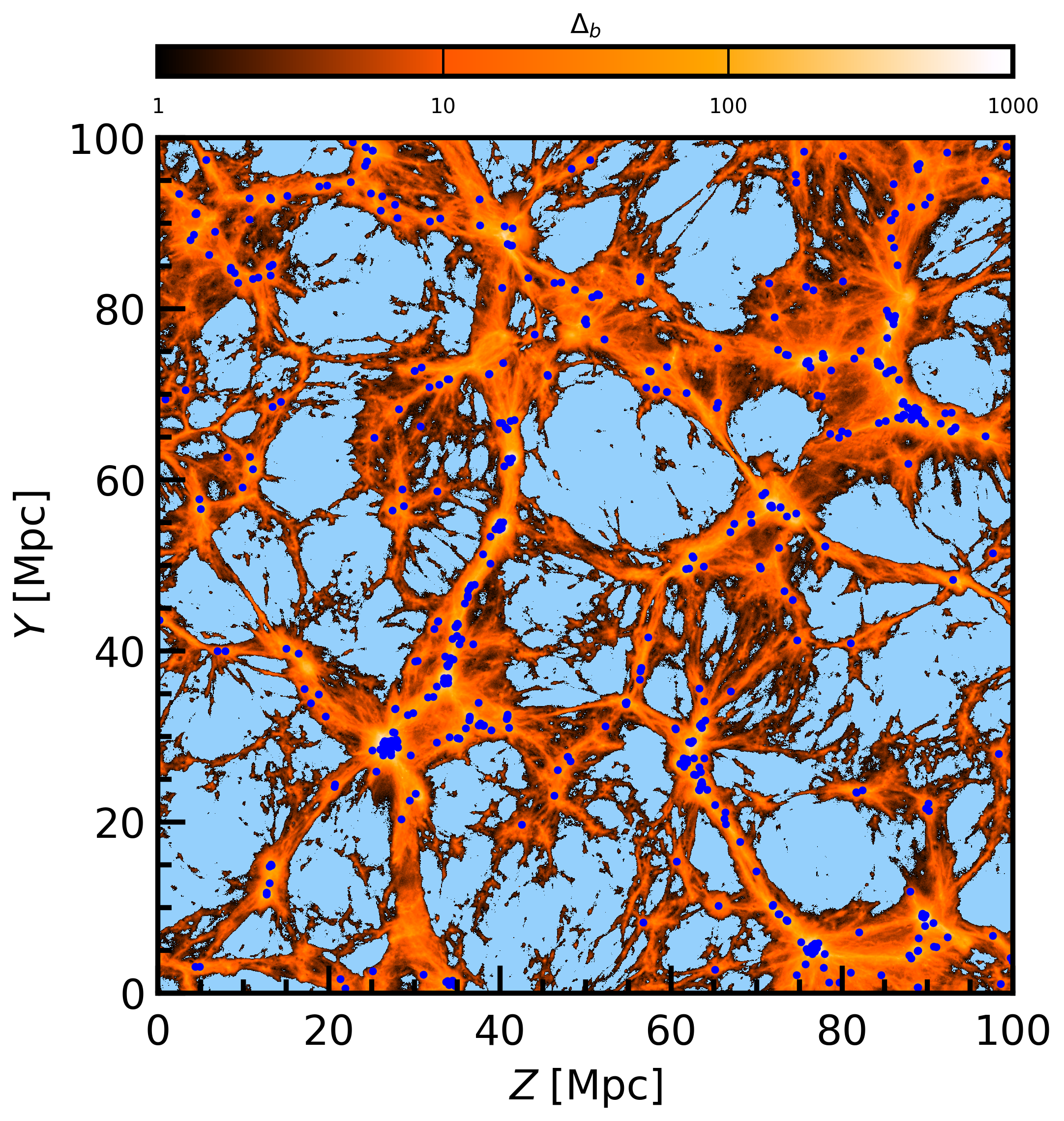}
\caption{Projected diffuse baryon density within a representative 5 Mpc slice (colour map) above the mean baryon density and galaxies brighter than M$_r$ = -18.4 (blue dots) in the EAGLE simulation. The light blue colour denotes the regions below the mean baryon density. The density was computed by co-adding the masses of individual gas particles and dividing by the projected volume in each pixel.}
\label{full_slice_hr.fig}
\end{figure*}

In order to obtain a general overview of the large scale distribution of the EAGLE baryons, we first computed the baryon densities within a representative 5 Mpc thick slice in a grid with a cell size of 0.1 Mpc. We subsequently projected the densities into a plane to obtain a 2D density distribution (see Fig.~\ref{full_slice_hr.fig}).
Consistently with earlier simulation works,
the distribution indicates the large underdense voids and the filamentary Cosmic Web at overdensities of $\delta \sim$ 10--100. 

Throughout this work all the 2D projections were computed using the aforementioned grid. Unless otherwise stated, all values, distributions and profiles have been computed directly from the particle data.

\subsection{Galaxies}
\label{gal}
We utilised the EAGLE galaxy data \citep{mcalpine_helly_etal_2016} for experimenting with the filament detection (see Section~\ref{intro}).  For each galaxy we selected the coordinates, the dust corrected {\it r} band magnitudes and the virial radius R$_{200}$ of the FoF group the galaxy belongs to (the radius where the mean internal density is 200 times over the critical density of the Universe). Within the virial radius the galaxies contain a stellar mass which constitutes $\approx$ 3.4\% of the total baryon budget in the EAGLE simulation.

\begin{figure}
\includegraphics[width=9cm,angle=0]{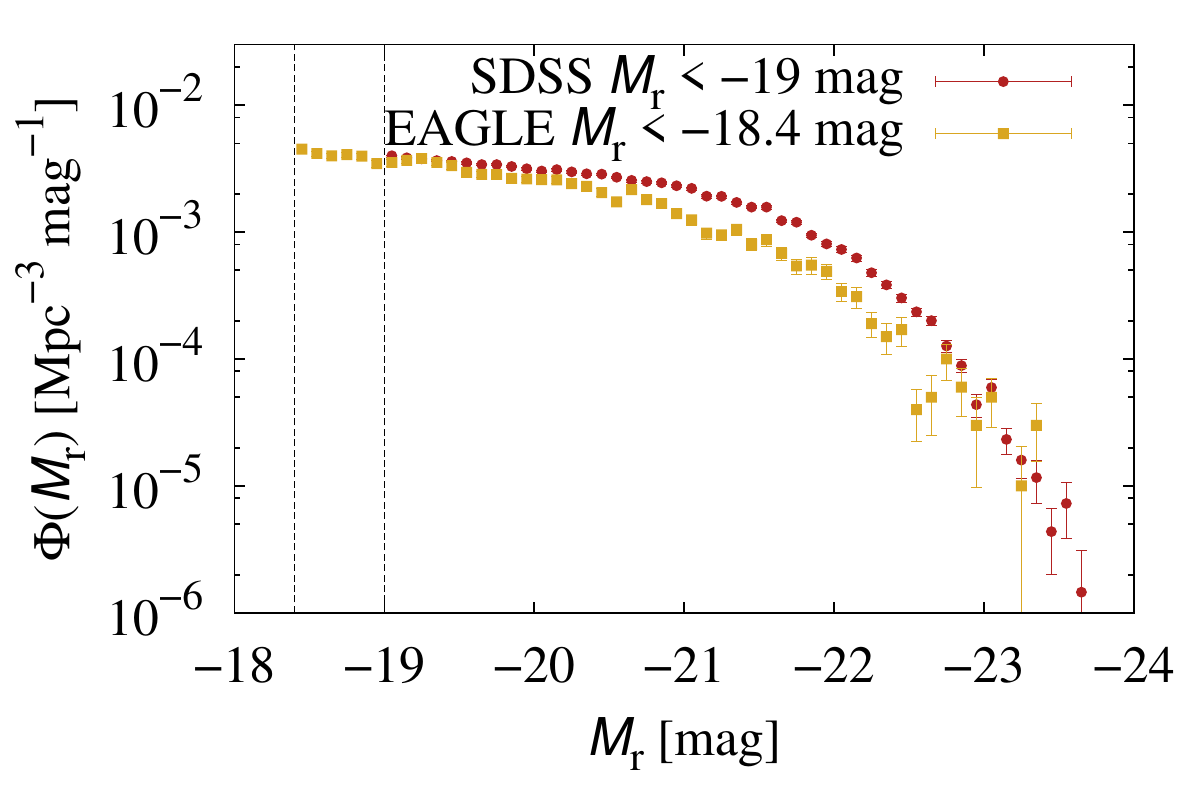}
\caption{Luminosity distributions of EAGLE (yellow points and error bars) and SDSS (red points and error bars) galaxies. The vertical dashed lines indicate the selected magnitude cuts of -18.4 and -19 for EAGLE and SDSS galaxies, respectively.}
\label{teet.fig}
\end{figure}

\begin{figure*}
\includegraphics[width=18cm,angle=0]{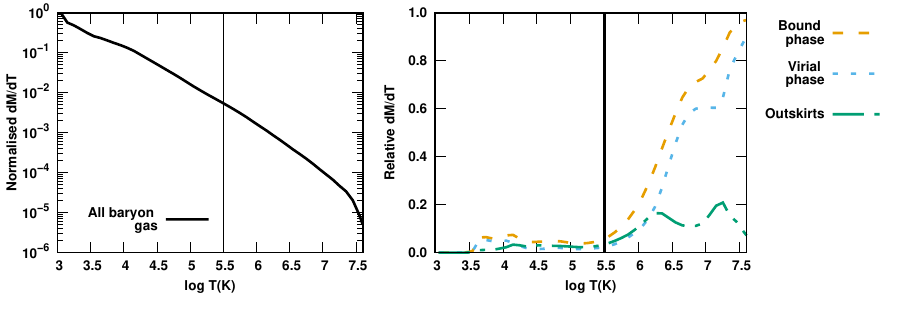}
\caption{
  {\it Left panel:} Distribution of the full EAGLE baryon gas mass content as a function of the temperature, normalised to unity at $\log{T(K)}~=~3$. {\it Right panel:} Corresponding distributions of different components, relative to the full distribution: the bound phase (orange dashed line), the virial phase within R$_{200}$ (blue dotted line) and the gas on the outskirts of galaxies and groups of galaxies (outside R$_{200}$ but within FoF halo, green dash-dotted line). The black vertical line indicates the approximate lower limit for the hot WHIM, i.e. $\log{T(K)}~=~5.5$.
} 
\label{mass_temperature.fig}
\end{figure*}

Since our aim is to provide observationally relevant results, we must consider that EAGLE is volume-limited while the observational surveys are typically
flux-limited. Our particular aim in the near future is to apply the results of this work to the Sloan Digital Sky Survey (SDSS, \citet{2014ApJS..211...17A}) up to redshift z $= 0.05$, since our adopted Bisous filament detection algorithm 
works well up to this distance \citep{2014MNRAS.438.3465T}.
At z $= 0.05$, the limiting SDSS magnitude corresponds to M$_{r} \approx -19$.
Unfortunately the luminosity distribution in EAGLE is somewhat different from SDSS (see Fig.~\ref{teet.fig}). Namely, the EAGLE distribution decreases faster than that of SDSS with higher luminosities. The behaviour is similar as when comparing the EAGLE simulations with the GAMA data at z $= 0.1$  \citep{2015MNRAS.452.2879T}. Thus, the 
EAGLE galaxy density is smaller than that of SDSS, if using the same magnitude limit. However, the galaxy density is the primary parameter determining the performance of our filament
detection algorithm (M. Muru, priv. comm.). Thus, to approximate the SDSS survey at z~$\le$~0.05, we have to include fainter EAGLE galaxies by using M$_r$ = -18.4 to match the SDSS M$_r$ = -19 galaxy density.
With this fainter cut we obtained a sample of filaments for the EAGLE simulation with similar statistical properties as those of the SDSS filaments. We assume that the slight mismatch of the EAGLE and SDSS magnitude limits does not produce complications in our analysis since the spatial distributions of the non-overlapping populations are expected to be similar: the galaxies are mostly in groups. A 2 Mpc smoothing length when computing the luminosity density field spreads the entire group luminosity into the IGM regions that are of interest in this work (see Section \ref{classification}).

Visual inspection indicates that the galaxy distribution follows the filamentary diffuse baryon distribution well (see Fig.~\ref{full_slice_hr.fig}).
This is consistent with \citet{2015A&A...583A.142N}, who reported a correlation between the density of the WHIM and the galaxy luminosity density in another simulation. This behaviour is consistent with the role of the dark matter in our adopted scenario, whereby the dominating dark matter gravity at the filaments enhances the baryon and galaxy density.

\section{The missing baryons}

\subsection{The problem}
\label{problem}
\citet{2012ApJ...759...23S} presented a popular census of the observational status of the cosmic baryon budget at low redshift, which indicated that $\sim 30\pm10$\% of the baryons predicted by the concordance cosmology have not been detected.
Later developments in the FUV measurements of AGN with HST/COS \citep{2016ApJ...817..111D} have refined the above account.
The \ion{H}{I}-traced phase (Lyman-Alpha forest and broad Lyman-Alpha absorbers (BLAs)) constitutes 25$\pm 5$\% of the baryon budget at z $=0.0-0.5$ (see Table~\ref{baryon_budget.tab}), which is smaller than estimated in \citet{2012ApJ...759...23S} (42$\pm$13\%). Assuming that the  observational status of other significant baryon components
(i.e. stellar mass and CGM in galaxies and the diffuse intergalactic baryons within galaxy groups and clusters)
is as estimated in \citet{2012ApJ...759...23S}, the \citet{2016ApJ...817..111D} estimate of the Lyman-Alpha (Ly$\alpha$) abundance yields a directly observational baryon budget of $\approx$ 40\% of the expected cosmic density.

In addition to the above directly observational cosmic hydrogen, there is also observational evidence of warm-hot $\log{T(K)} \approx 5.0-5.5$ cosmic baryons via extragalactic $\ion{O}{VI}$ absorbers \citep[e.g.][]{2016ApJ...817..111D}.
In the above work the cosmic contribution of the baryons traced by \ion{O}{VI} is reported as 11$\pm$1\%. There are several uncertainties with the usage of this result in the cosmic baryon budget.

First, the above conclusion requires information on the temperature and metal abundance of the gas so that the \ion{O}{VI} measurements can be converted into the underlying dominating baryon component, the hydrogen mass. This information is not attainable via \ion{O}{VI} line measurements only.  The solution used in \citet{2016ApJ...817..111D} (and in similar earlier works by \citealt{2012ApJ...759..112T} and \citealt{2012ApJ...759...23S}) is to utilise the correlation between the ion fraction, metal abundance and the ion column density from \textit{simulations}. This complicates the interpretation of the derived hydrogen population as \textit{observed} cosmic baryons.  

Second, the reported $\approx$ 10\% statistical precision indicates that the uncertainties of the temperatures and metallicities are relatively small. Observational estimates do not exist and the 
scatter of the correlation due to cosmic variation, available in the simulations,  is not propagated to the uncertainties.  Thus, the reported uncertainties are underestimated.

Third, \ion{O}{VI} traces hydrogen within the same temperature range as the BLA measurements, i.e. $\log{T(K)} = 5.0-5.5$. If the \ion{O}{VI}-traced baryons and those directly detected via BLA are treated independently, we double-count the overlapping, uncertain population.

Thus, the \ion{O}{VI}-traced contribution lies between 0\% (full overlap with the Ly$\alpha$ observations) and $\approx 10\%$ (completely independent component), yielding a total missing baryon fraction within the range of $\approx 50-60\%$.  This would be a significant overestimate if the true  \ion{O}{VI} - traced cosmic baryon density is much higher than the reported 11\%, which is possible given the underestimated uncertainties.

\begin{table}
 \centering
  \caption{Estimates for the observational baryon budget. 
  \label{baryon_budget.tab}}
    \begin{tabular}{lcc}
  \hline\hline
                            &                                   &            \\ 
  component   	            & $\Omega_{b,i}$ / $\Omega_{b,planck}$ & reference  \\
                            &                                   &            \\
\hline 
                            &                                   &            \\
Ly$\alpha$\tablefootmark{a} &  25\%                       & D\tablefootmark{b}         \\
FUV metals\tablefootmark{c} &  0-11\%                       & D          \\
stars\tablefootmark{d}      &  7\%                        & S\tablefootmark{e}     \\
ICM                         &  4\%                      & S          \\ 
CGM                         &  5\%                        & S          \\ 
&                                   &            \\
Total                       &  41-52\%                      & S\&D         \\
\hline                  
                            &                                   &            \\

\end{tabular}
\tablefoot{\\
\tablefoottext{a}{Including BLAs (in FUV).} 
\tablefoottext{b}{\citet{2016ApJ...817..111D}}
\tablefoottext{c}{ Tracing $\log{T(K)} = 5-5.5$ gas not observed via BLAs.} 
\tablefoottext{d}{Stellar mass in galaxies.} 
\tablefoottext{e}{\cite{2012ApJ...759...23S}.}
}
\end{table}

\begin{table}
 \centering
  \caption{Hot gas ($\log{T(K)}$ > 5.5) fraction of the total baryon content within the EAGLE simulation. 
  \label{eagle_budget.tab}}
    \begin{tabular}{lcc}
  \hline\hline
                                   &                                   &           \\ 
  component   	                   & $\Omega_{b,i}$ / $\Omega_{b,planck}$ &          \\
                                   &                                    &          \\
\hline 
                                   &                                    &          \\
Total                              &                    42\%            &          \\
                                   &                                    &          \\
Within R$_{200}$                    &                    13\%            &          \\
Hot WHIM \tablefootmark{a}         &                    29\%            &          \\
                                   &                                    &          \\
Outskirts\tablefootmark{b}         &                    5\%             &          \\
Bisous filaments \tablefootmark{c} &           25\%                     &          \\
NEXUS+ filaments \tablefootmark{c} &           23\%                     &          \\
\hline 
\end{tabular}
\tablefoot{\\
\tablefoottext{a}{ ``Total'' - ``Within R$_{200}$''.}
\tablefoottext{b}{Gas outside R$_{200}$ but within FoF halo.}
\tablefoottext{c}{Outside R$_{200}$, including outskirts.}
}
\end{table}

\subsection{Our approach}
In this work we utilise the hypothesis that  a large fraction of the missing baryons resides in the poorly observed hot gas phase.
Assuming collisional ionisation equilibrium, the ionisation fraction of the most important FUV ion \ion{O}{VI} peaks at $\log{T(K)}\approx 5.5$ and decreases rapidly with higher temperatures 
where the most important X-ray ion \ion{O}{VII} dominates. The transition is not clear-cut, as shown by \citet{2018MNRAS.477.450N} in the overlap of temperature-dependent \ion{O}{VI} and \ion{O}{VII} ion mass distributions around $\log{T(K)}=5.5$ at typical filament baryon overdensities of 10-100 in IllustrisTNG simulations. In order to minimise the robustly detected FUV phase \citep{2016ApJ...817..111D} in our work we adopt $\log{T(K)} \ge 5.5$ as one criterion for the missing baryons. $\approx$ 42\% of the EAGLE baryon content is in the form of gas above this temperature limit (see Table \ref{eagle_budget.tab}).

\subsubsection{ The bound phase}
\label{bound}
Previous simulations \citep[e.g.][]{2001ApJ...552..473D, 2006MNRAS.370..656D,2009ApJ...697..328B,2012MNRAS.425.1640T,2012MNRAS.423.2279C, 2018MNRAS.473...68C, 2019MNRAS.486.3766M} have indicated that a significant hot baryon reservoir is located within the filaments of the Cosmic Web.
We utilise the results of the Friends-of-Friends (FoF) procedure to separate the more easily detectable hot and dense phase bound to galaxies and galaxy groups\footnote{Galaxy clusters are largely missing due to the small volume covered by EAGLE ($100^3 \, \mathrm{Mpc}^{3}$).}  from the purely intergalactic phase as follows.

The FoF procedure has been applied by the EAGLE team to the dark matter particle distribution to define through constant density boundaries the dark matter haloes.
A gas particle whose neighbouring dark matter particle belongs to a FoF halo has been defined as a halo particle. 
We assume that such particles form the bound phase of baryons.
We separate the bound phase into two environments: 1) the virial phase, i.e. gas particles located within the virial radius R$_{200}$ of a FoF halo  given in the public EAGLE database \citep{mcalpine_helly_etal_2016} and
2) the outskirts phase, i.e. gas particles occupying the FoF halo outside  R$_{200}$.

In order to understand the baryon properties of the bound phase, we examined the mass distribution of the gas within FoF haloes as a function of the temperature.  
We divided the temperature range of $\log{T(K)}~=~3-8$ into 50 bins of $\log{T(K)}~=~0.1$ width. We then added the mass of gas particles whose temperatures were within the boundaries of a given bin. Subsequently we divided the resulting masses with the temperature bin size, thus obtaining the distribution dM/dT (T). We normalised the distribution to the total gas mass (see Fig.~\ref{mass_temperature.fig}).
We repeated the process separately for the gas within the virial radius and the gas in the outskirts of galaxies and groups of galaxies to compare the different environments. 

The temperature distribution of the bound phase indicates that its contribution to the total baryon budget at 
$\log{T(K)}~\le~6$ is smaller than 10\% (see Fig.~\ref{mass_temperature.fig}, right panel).  
Towards higher temperatures the relative importance of the bound phase increases: at $\log{T(K)}~>~6.5$ the bound phase contains $\approx$ 75\% of the EAGLE baryon mass.
We assume that this is due to the gravitational collapse shocks involved in heating the bound baryons being more energetic than the accretion shocks heating the intergalactic baryons.
However, since the absolute amount of baryons decreases rapidly towards the highest temperatures (see Fig.~\ref{mass_temperature.fig}, left panel), the contribution of the bound phase to the baryons in the missing baryon temperature range ($\log{T(K)}~>~5.5$) is only $\approx$~44\%. Thus, there is ample room for a large fraction of the missing baryons to reside in other environments (particularly in filaments, see Section~\ref{fila}).

At the low end of the missing baryon temperature range ($\log{T(K)}~=~5.5-6.0$) the outskirts contribute $\approx$ 50\% to the bound phase (see Fig.~\ref{mass_temperature.fig}).
At higher temperatures the gas within R$_{200}$ dominates the bound phase. Thus, according to the EAGLE simulation, the contributions of the outskirts and R$_{200}$ to the baryon budget in the missing baryon temperature range ($\log{T(K)}~>~5.5$) are $\approx$ 12\% and $\approx$ 32\%, respectively. Their respective contributions to the full EAGLE baryon budget are $\approx$ 5\% and $\approx$ 13\% (see Table \ref{eagle_budget.tab}). Stars within R$_{200}$ contribute an additional $\approx$ 3\% to the total baryon mass.
In our filament analysis below, we include the hot baryons of the outskirt regions in our missing baryon sample i.e. we exclude only the baryons residing within R$_{200}$.

\begin{figure*}
\vbox{
\hbox{
\includegraphics[width=9cm,angle=0]{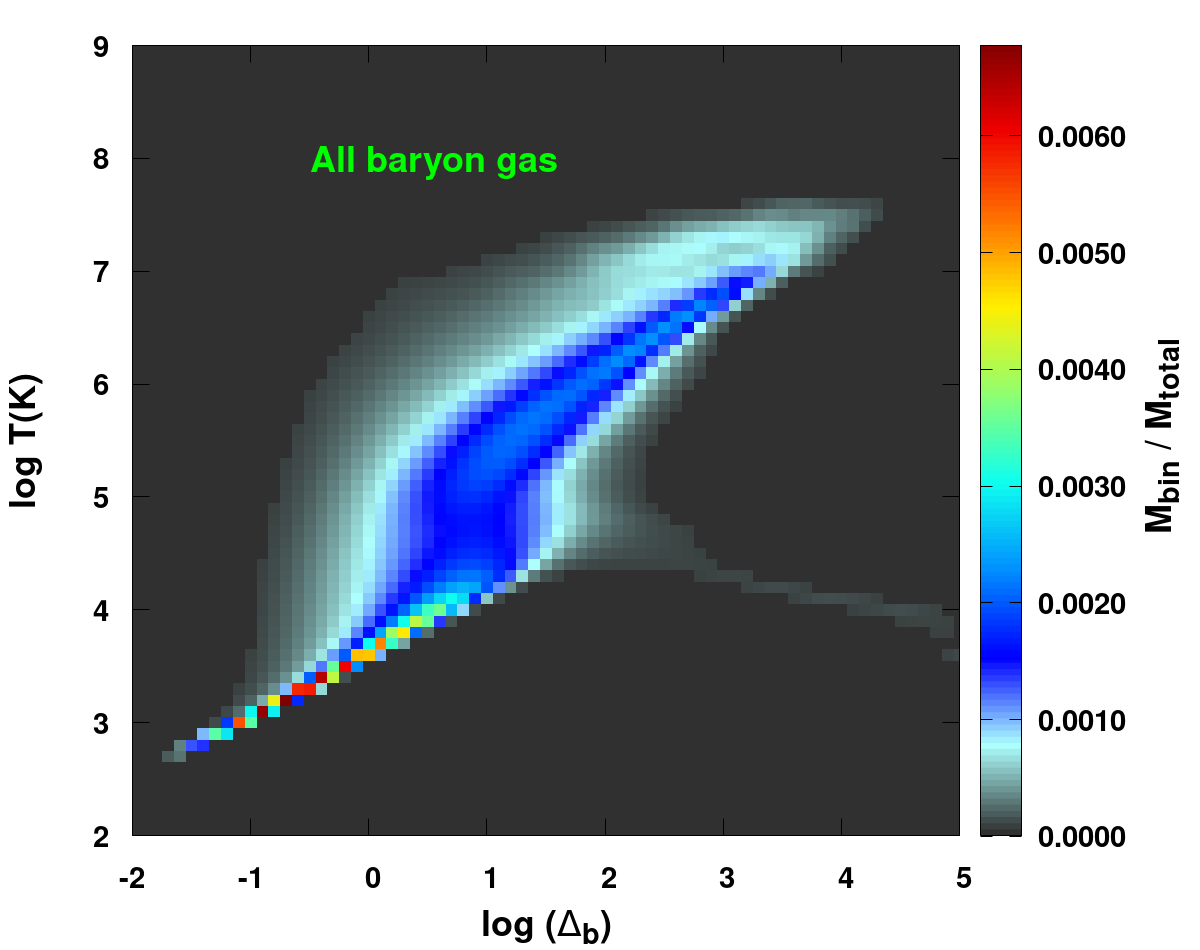}
\includegraphics[width=9cm,angle=0]{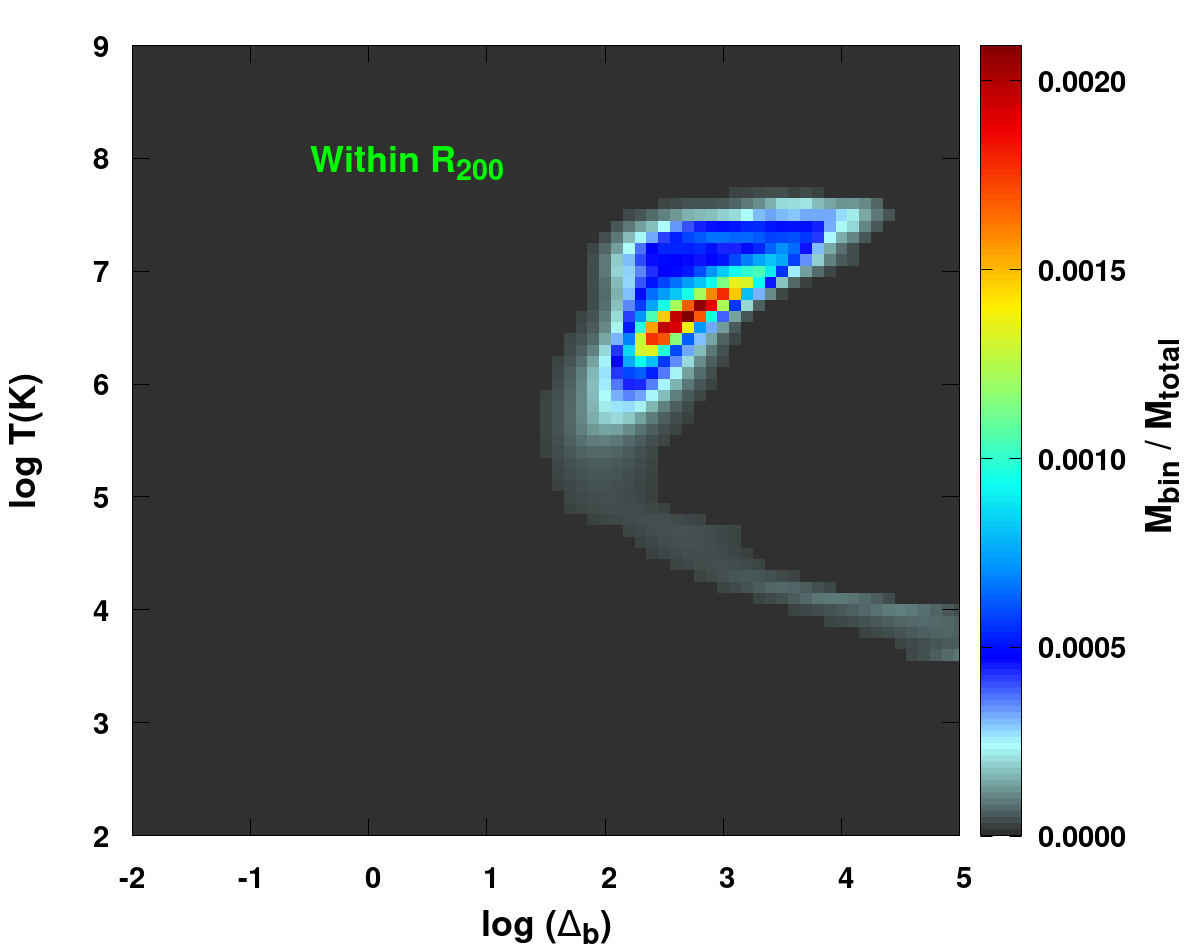}}
\hbox{
\includegraphics[width=9cm,angle=0]{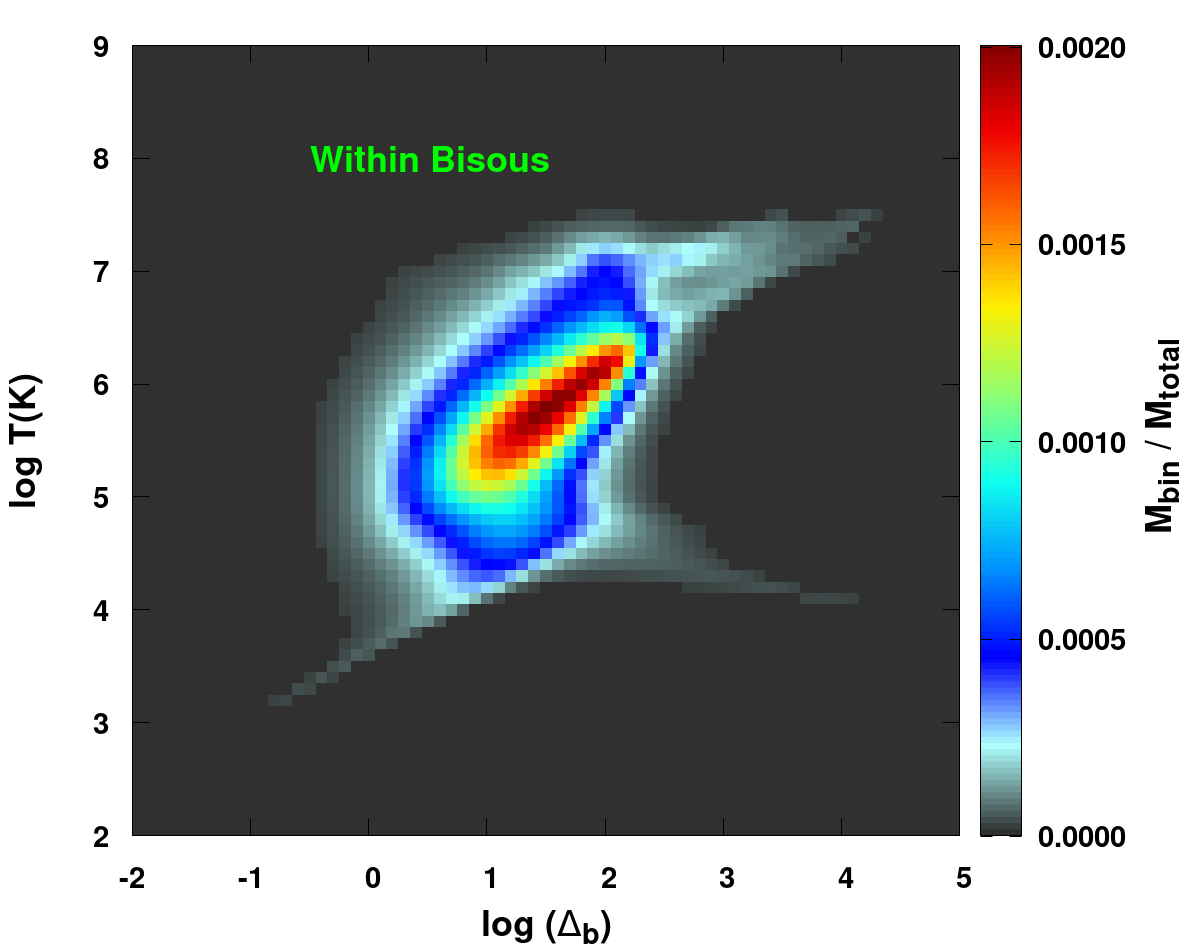}
\includegraphics[width=8cm,angle=0]{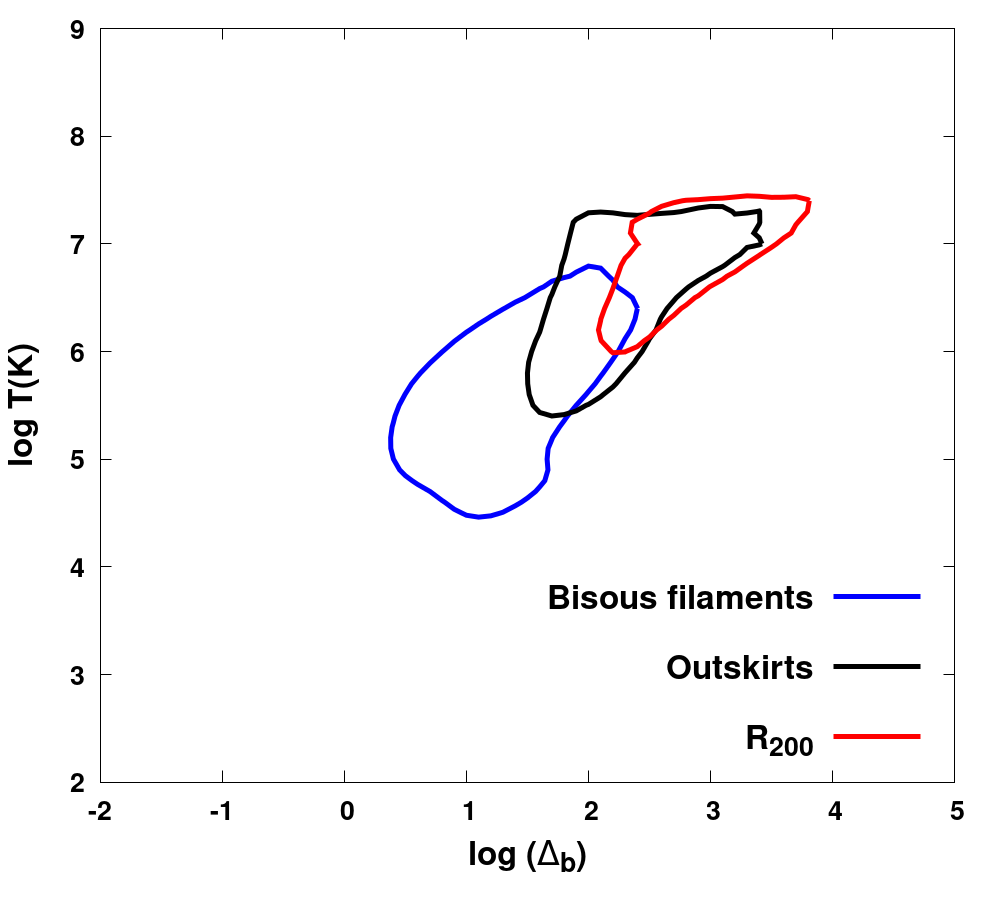}}
}
\caption{Baryon density contrast - temperature phase diagrams (for the convenience of plotting we use the density contrast $\Delta_{b} = 1 + \delta_{b}$ rather than the overdensity $\delta_{b}$). All the gas in the full EAGLE volume is shown in the top left panel, the gas within R$_{200}$ in the top right panel and the gas within the Bisous filaments excluding gas within R$_{200}$ in the bottom left panel. For each temperature and density bin we co-added the mass of particles and divided by the total mass of the full box. We cut the colour scale at 1\% level of the maximum value for clarity. The bottom right panel indicates the isomass curves containing 68\% of the baryon mass in Bisous filaments excluding R$_{200}$ (blue line), within the outskirts (black line) and within R$_{200}$ (red line).  The low-density, low-temperature track seen in the upper left panel corresponds to the photo-heated IGM.
  }
\label{phasediag.fig}
\end{figure*}

To understand the bound phase better, we produced temperature-density phase diagrams as follows. We binned the density contrast and temperature into 70 bins each (see Fig. \ref{phasediag.fig}). The range for the overdensities was $\Delta_{b}$ = 0.01-$10^{5}$ (bin width of $\log{\Delta_{b}} = 0.1$) and for the temperature $\log{T(K)}=2-9$ (bin width of $\log{T(K)} = 0.1$). Then, we summed the masses of the particles that fell within the ranges of the bins. Finally, the mass in each bin was divided by the total mass of all the gas particles to calculate the corresponding mass fractions. In addition to the full EAGLE gas content, we performed the above analysis for particles within R$_{200}$ and outskirts separately.

The temperature-density phase diagram of the bound phase indicates the expected behaviour that the denser gas is hotter (see Fig.~\ref{phasediag.fig}). 
The concentration of virial gas above $\log{T(K)}~=~6$  (discussed above) corresponds to a density concentration above $\Delta_{b} \sim 100$. This phase is relatively compact and dominates the hottest and densest domain. 
The outskirt gas has similar behaviour but at a bit lower temperatures and densities (see Fig.~\ref{phasediag.fig}).
The outskirts gas phase matches the dense, high-temperature part of the Bisous filament phase (see Section \ref{thermodyn} for more discussion).
A small fraction of the diffuse baryons (1.8\%) is locked within the central regions inside and around galaxies.
This low temperature - high density phase is seen as a low-level deviation from the main component of the virialised gas (see upper right panel of Fig.~\ref{phasediag.fig}).

\subsubsection{ The missing baryon sample in EAGLE}
\label{definition}
 The virial phase behaves smoothly in the temperature-density plane (see Fig.~\ref{phasediag.fig}) and dominates the hottest and densest domain. Thus, the exclusion of the virial phase
from the full EAGLE diffuse baryon population would provide a simple way of separating the more easily detectable high density and high temperature gas from the missing baryons.
After the additional low temperature ($\log{T(K)}~<~5.5$) cut, only $\approx$ 7\% of the total baryon mass is outside the canonical WHIM baryon density interval of 1-100 times the mean.
Similarly, only $\approx$ 2\% of the total baryon mass exceeds the canonical WHIM upper temperature limit of $\log{T(K)}$~=~7.

 While being simple, the exclusion of all the hot virial gas in EAGLE from the missing baryon category would yield a somewhat underestimated missing baryon fraction.
  Namely, assuming that the extrapolation of the observed hot gas mass in the central regions of galaxies to the virial radii \citep[e.g.][]{2018ApJ...862....3B} is accurate, most of the virial gas within galaxies is currently unobserved, i.e. missing. Assuming that 1) the observed CGM contribution of $\approx$ 5\% to the cosmic baryon budget \citep{2012ApJ...759...23S} accurately corresponds to the observed hot gas in galaxies
 discussed in \cite{2018ApJ...862....3B}, and that 2) EAGLE accurately describes the contribution of the hot gas within R$_{200}$ to the
 cosmic baryon density ($\approx$ 13\%, see Section~\ref{bound}), by excluding the gas within R$_{200}$ we would ignore a missing baryon component amounting to $\approx$ 8\% of the cosmic baryon budget.

However, our focus in this work is the intergalactic WHIM and thus we will not analyse further the possible missing baryons inside R$_{200}$ (for a thorough analysis of the CGM in EAGLE see \citealt{2020MNRAS.498..574W}).
For the purposes of this work, we adopt the following definition of a missing baryon:  
a) its temperature is higher than $10^{5.5}$ K and b) it is outside the virial radius (R$_{200}$) of collapsed structures
(i.e. we consider that the outskirts phase belongs to the missing baryons category). 

The baryons satisfying these criteria ({\bf P$_{true}$}) correspond to $\approx$ 29\% of the total baryon mass in the EAGLE simulation.
This substantial reservoir of baryons that are currently difficult to observe is a significant share of the 
missing baryon fraction of  $\approx$ 50-60\% estimated with the baryon census of \citet{2012ApJ...759...23S} and \citet{2016ApJ...817..111D} (see Section \ref{problem}).
This supports our assumption that the hot WHIM is a major component of the missing baryon population.
The currently non-observable baryon population within R$_{200}$ of galaxies, if amounting to $\approx$ 8\% of the cosmic baryon density (see above),
would increase the missing baryon fraction in the form of hot gas to $\approx$ 37\% according to EAGLE. While this level is lower than the formal missing baryon fraction census of $\approx$ 50-60\% 
(\citet{2012ApJ...759...23S} and \citet{2016ApJ...817..111D}), the missing baryon census might be significantly overestimated due to the possibly underestimated \ion{O}{VI} -traced cosmic baryon density (see Section \ref{problem}).

\begin{figure*}
\includegraphics[width=18cm,angle=0]{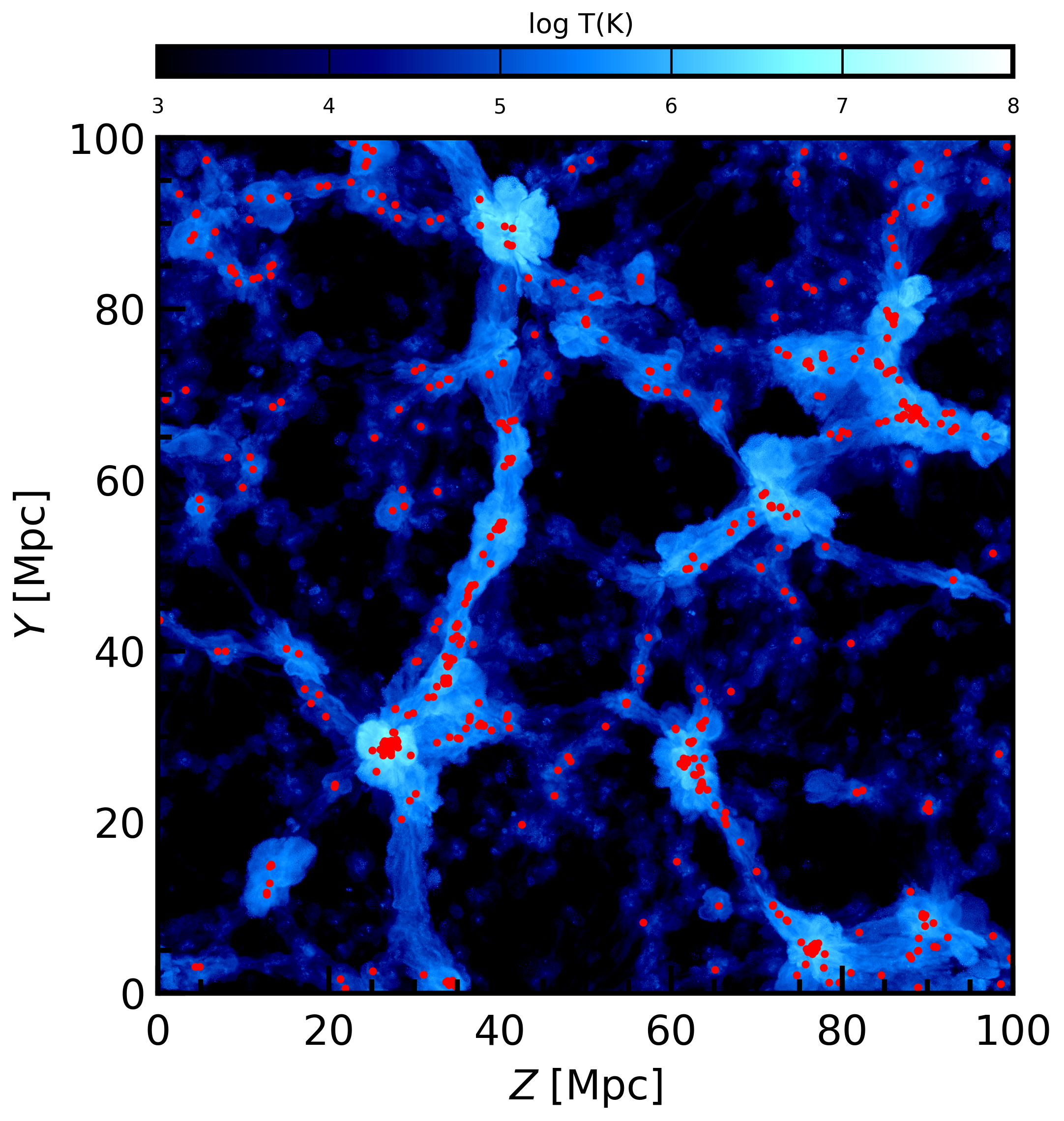}
\caption{Mass-weighted mean temperature distribution (in shades of blue) of the intergalactic gas, i.e. gas outside R$_{200}$, within the same slice as Fig. \ref{full_slice_hr.fig}. Red dots denote the location of galaxies brighter than  M$_r$ = -18.4.}
\label{tmap.fig}
\end{figure*}

\begin{figure*}
  \hspace*{1.7cm}
  \vbox{
\hbox{
\includegraphics[width=7.25cm,angle=0]{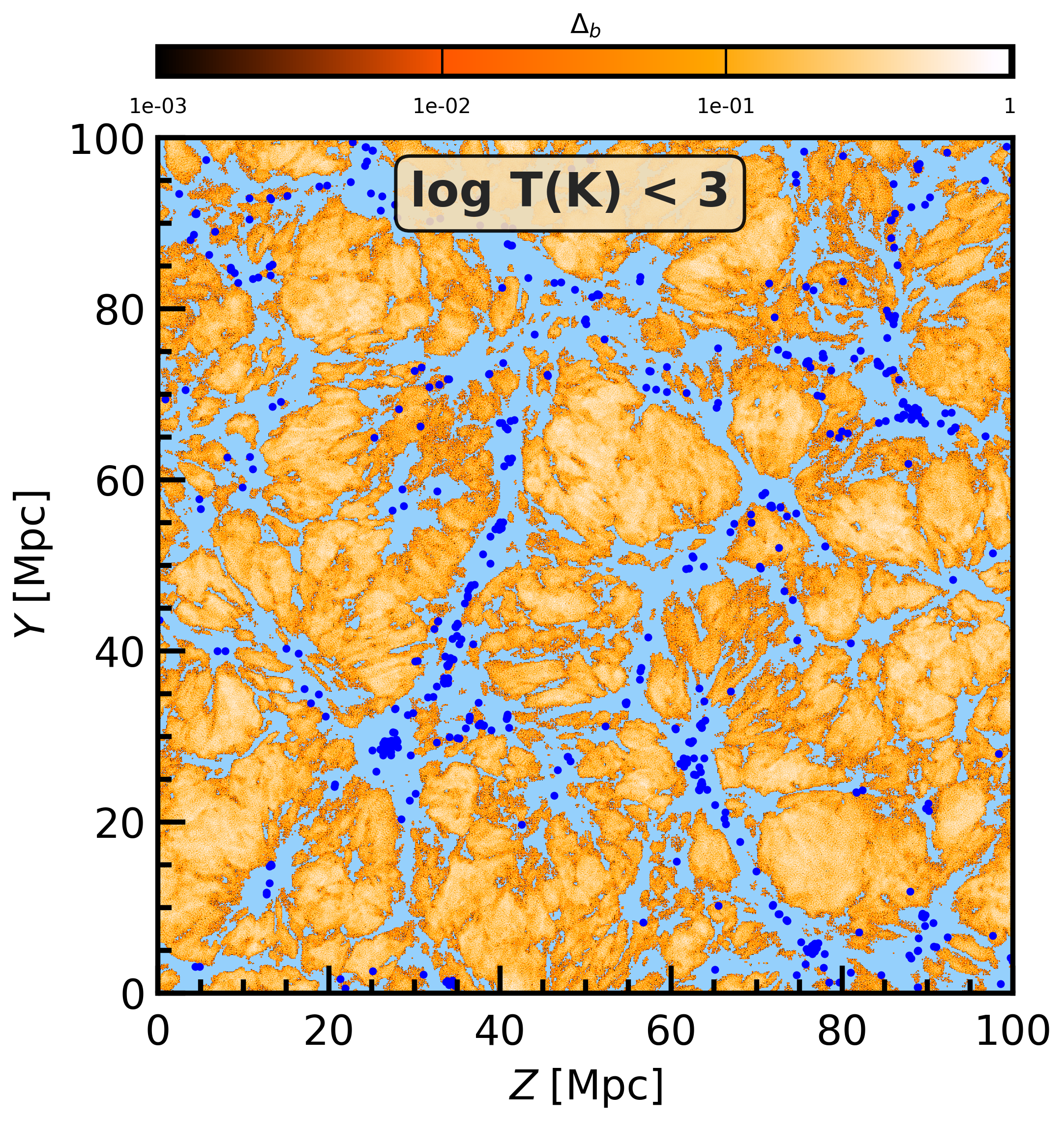}
\includegraphics[width=7.25cm,angle=0]{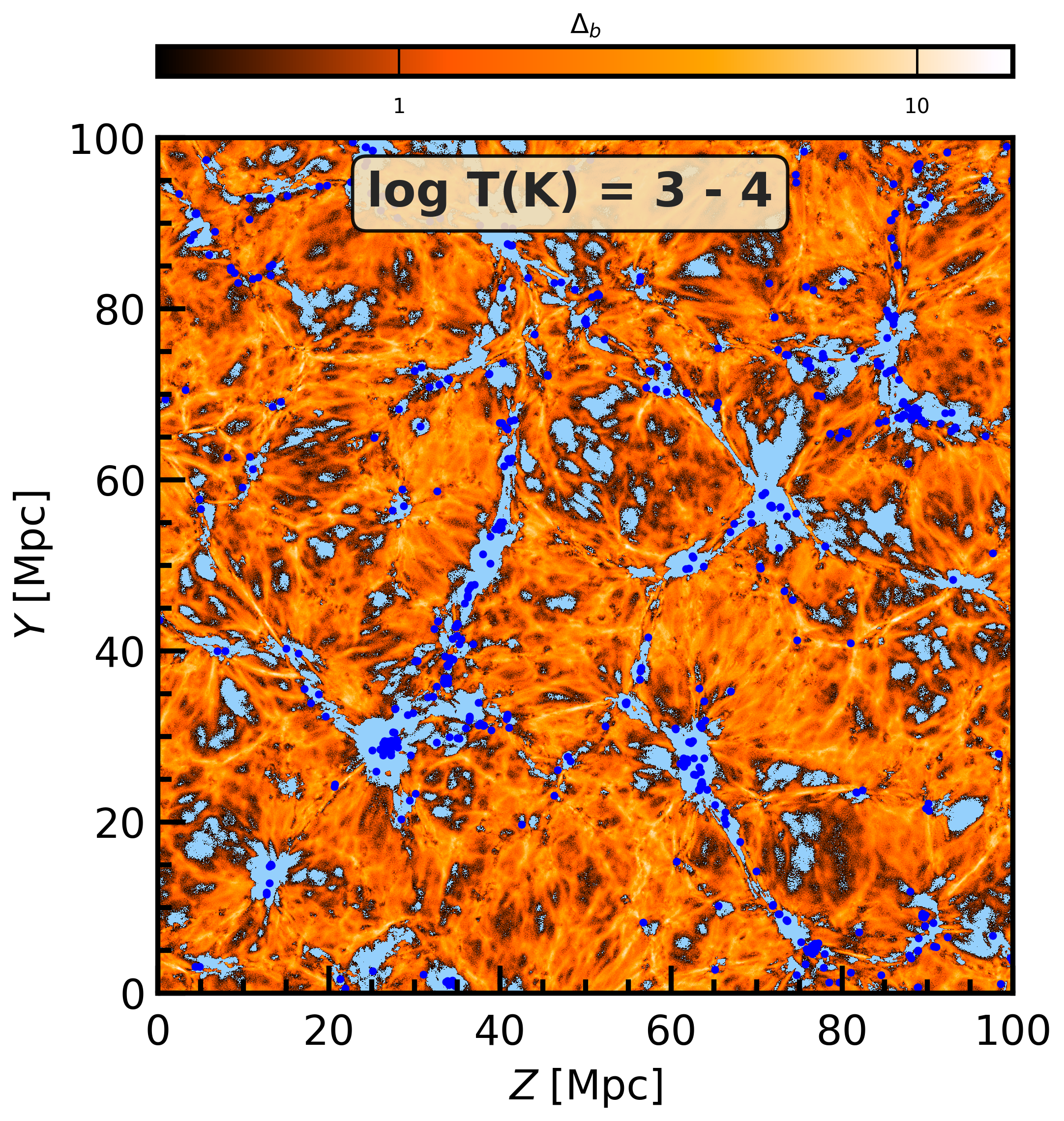}
}
\hbox{
\includegraphics[width=7.25cm,angle=0]{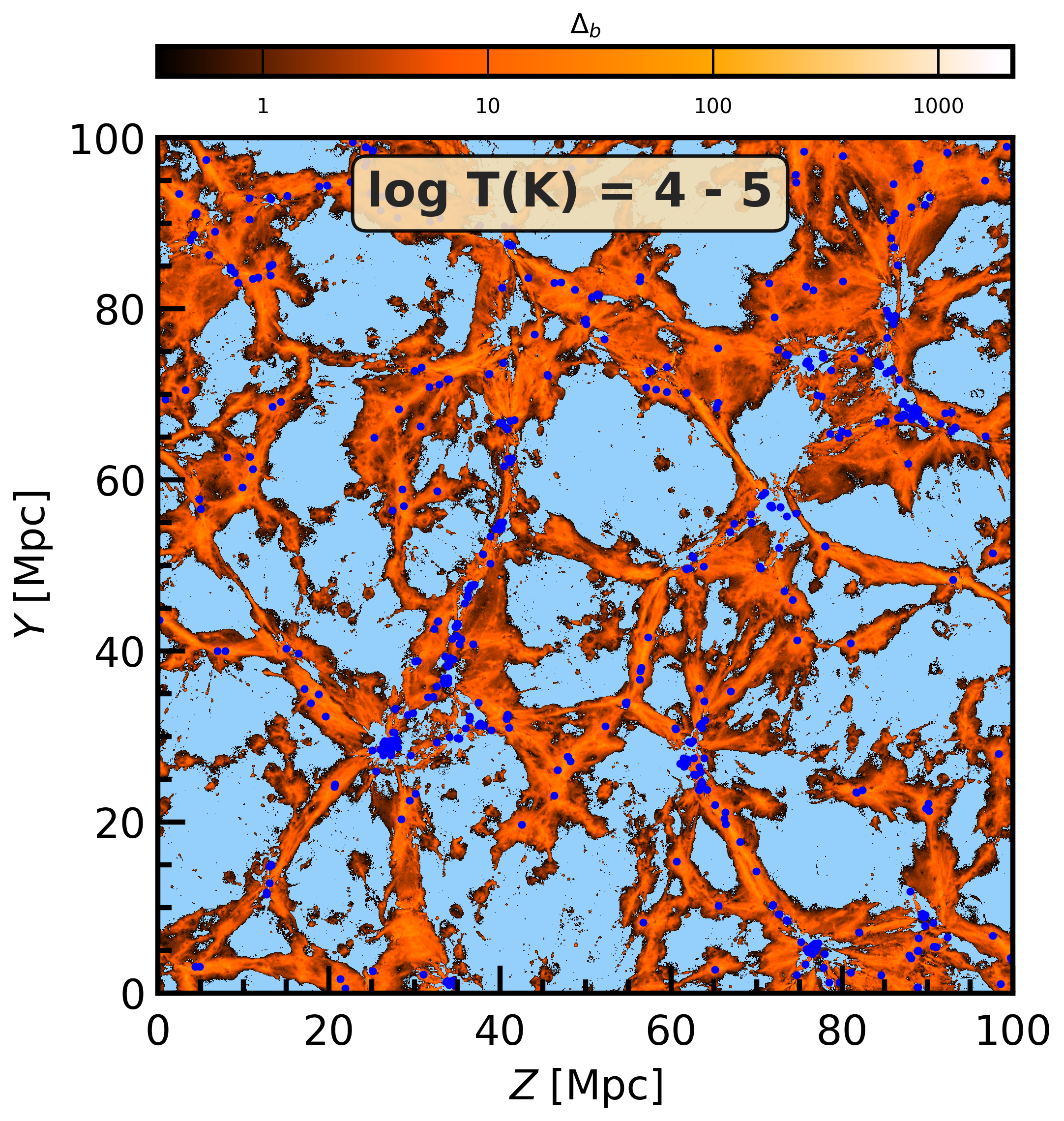}
\includegraphics[width=7.25cm,angle=0]{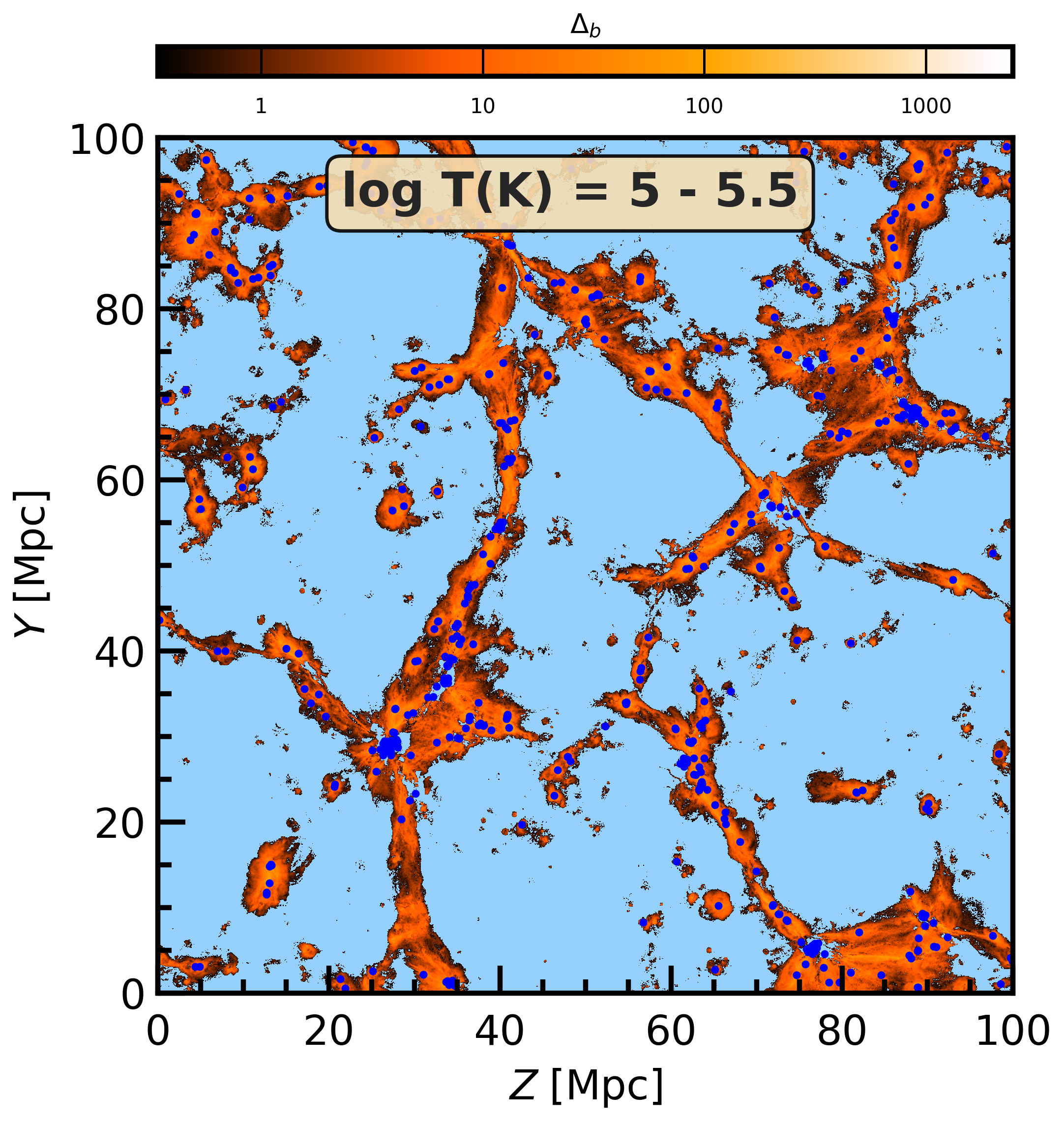}
}
\hbox{
\includegraphics[width=7.25cm,angle=0]{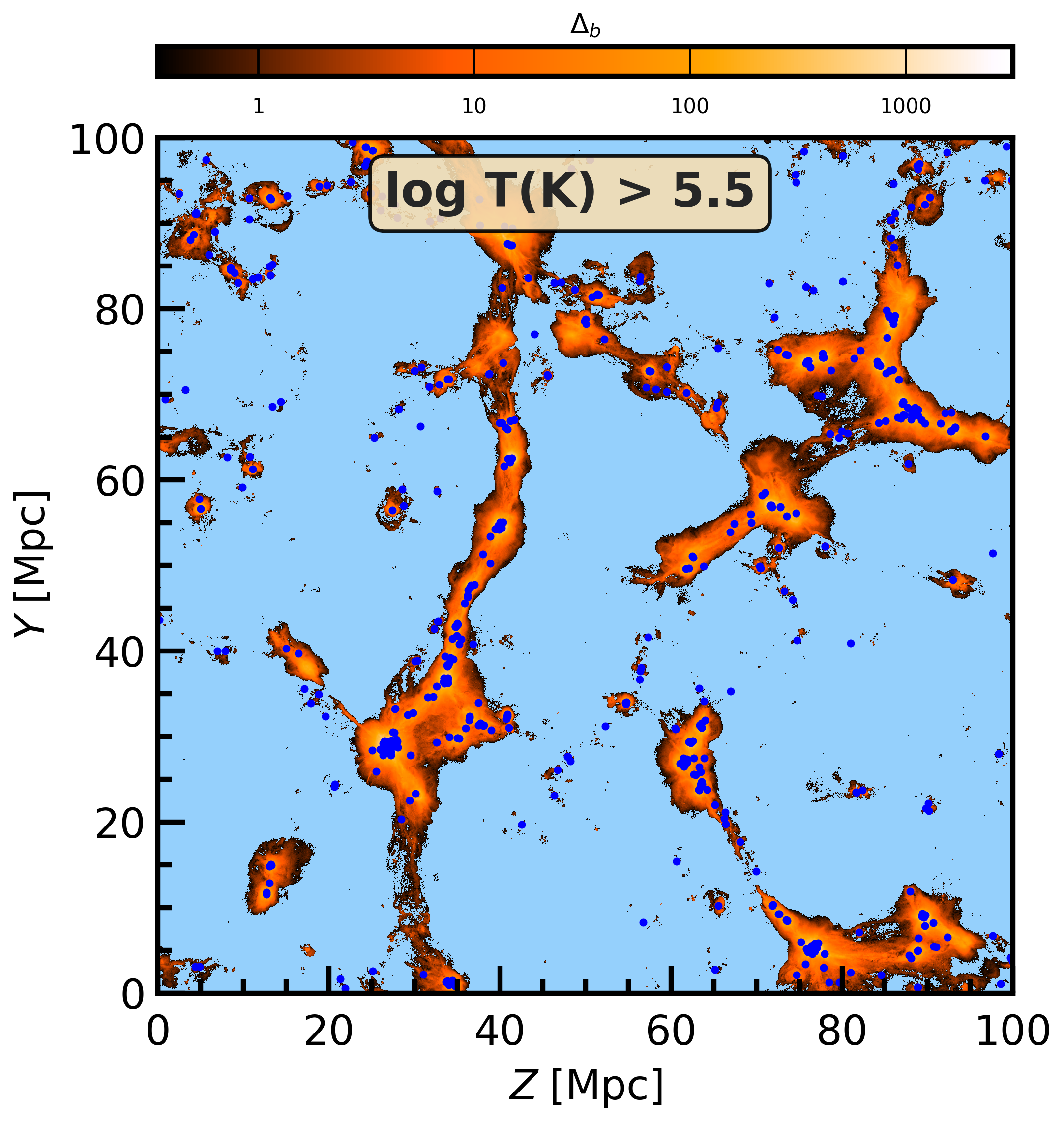}
\includegraphics[width=7.25cm,angle=0]{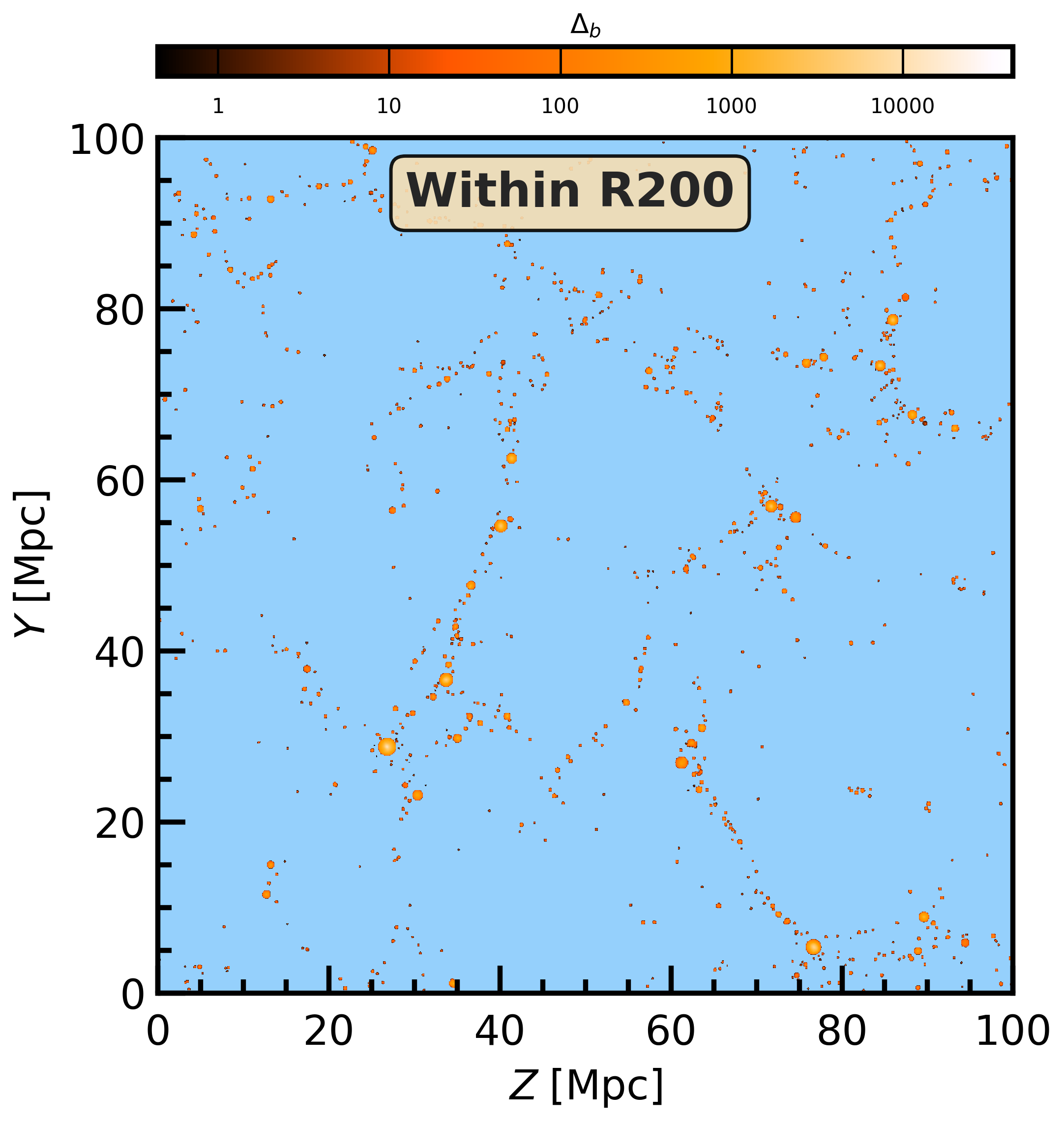}
}
}
  \caption{Projected diffuse baryon mass density contrast within the same 5 Mpc slice as Fig. \ref{full_slice_hr.fig}, but here in different temperature ranges.
    {\it Top left}: $\log{T(K)}~<~3$.
    {\it Top right}: $\log{T(K)}~=~3-4$.
    {\it Middle left}: $\log{T(K)}~=~4-5$.
    {\it Middle right}: $\log{T(K)}~=~5-5.5$ (the warm WHIM phase).
    {\it Bottom left}: $\log{T(K)}~>~5.5$ (the missing baryon phase).
    {\it Bottom right}: Gas within the virial radii R$_{200}$ of haloes, excluded from the other panels. 
  }
\label{phase_slices.fig}
\end{figure*}

\begin{figure*}
\vbox{
\hbox{
\includegraphics[width=9cm,angle=0]{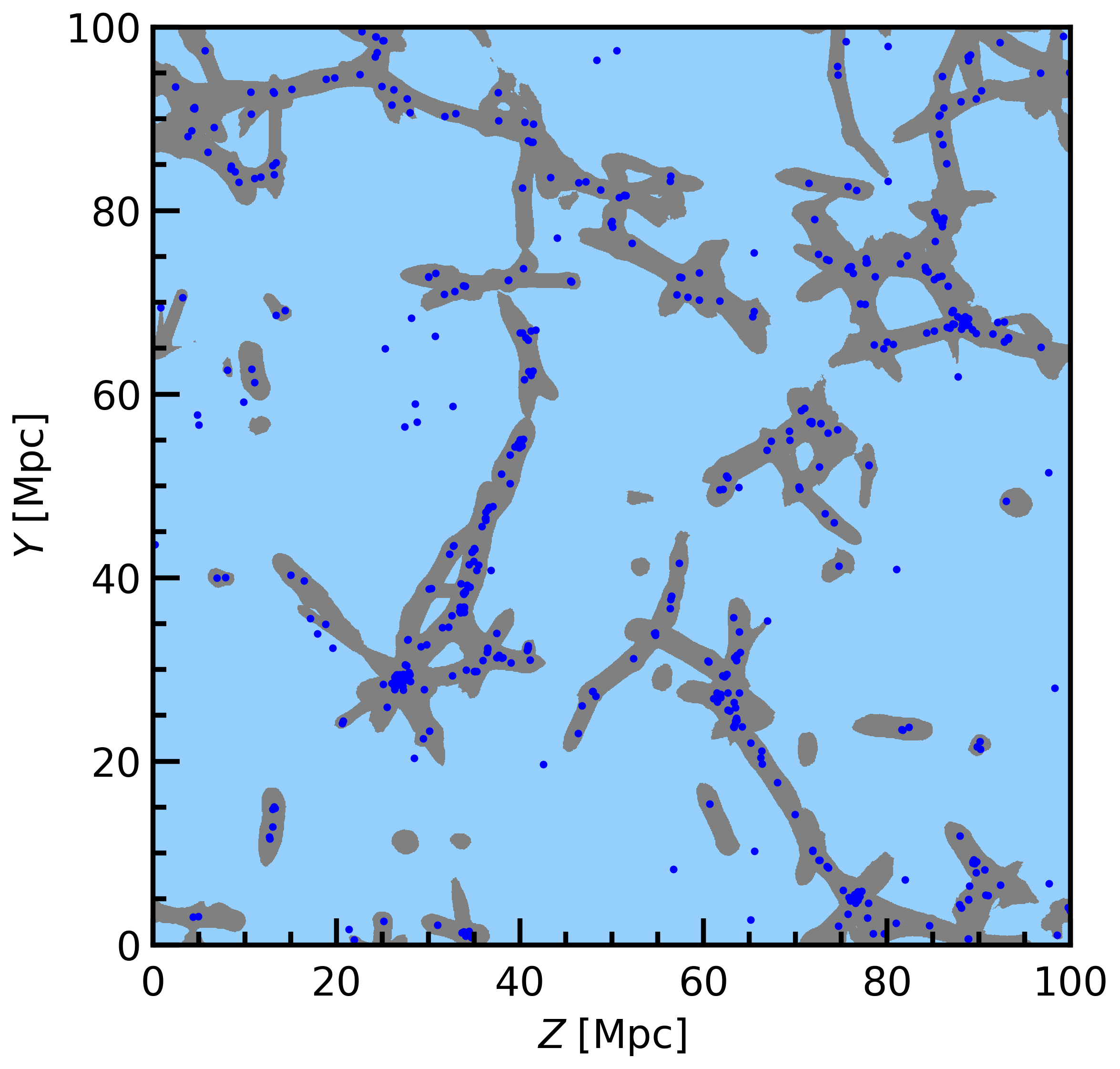}
\includegraphics[width=9cm,angle=0]{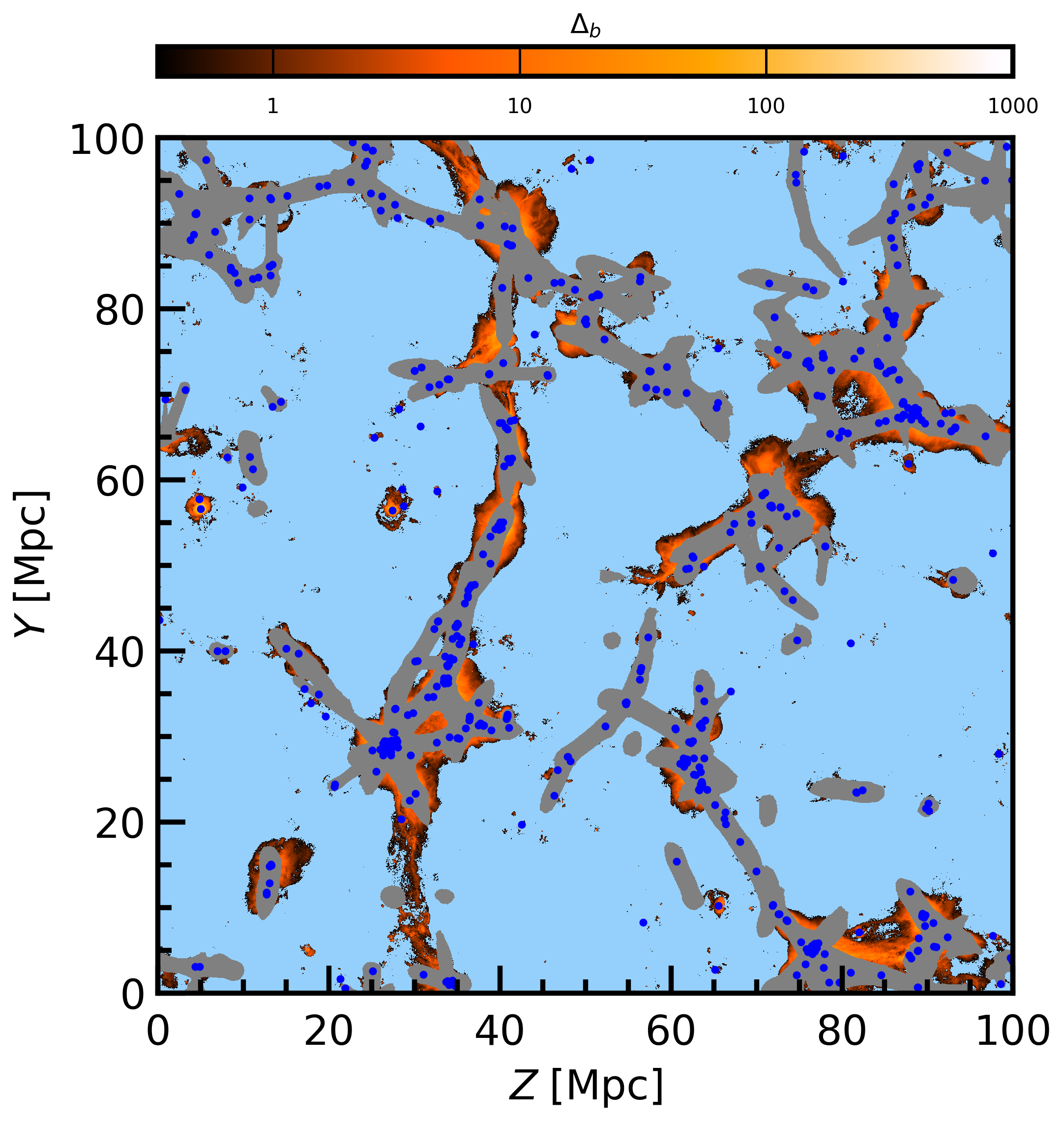}
}
\hbox{
\includegraphics[width=9cm,angle=0]{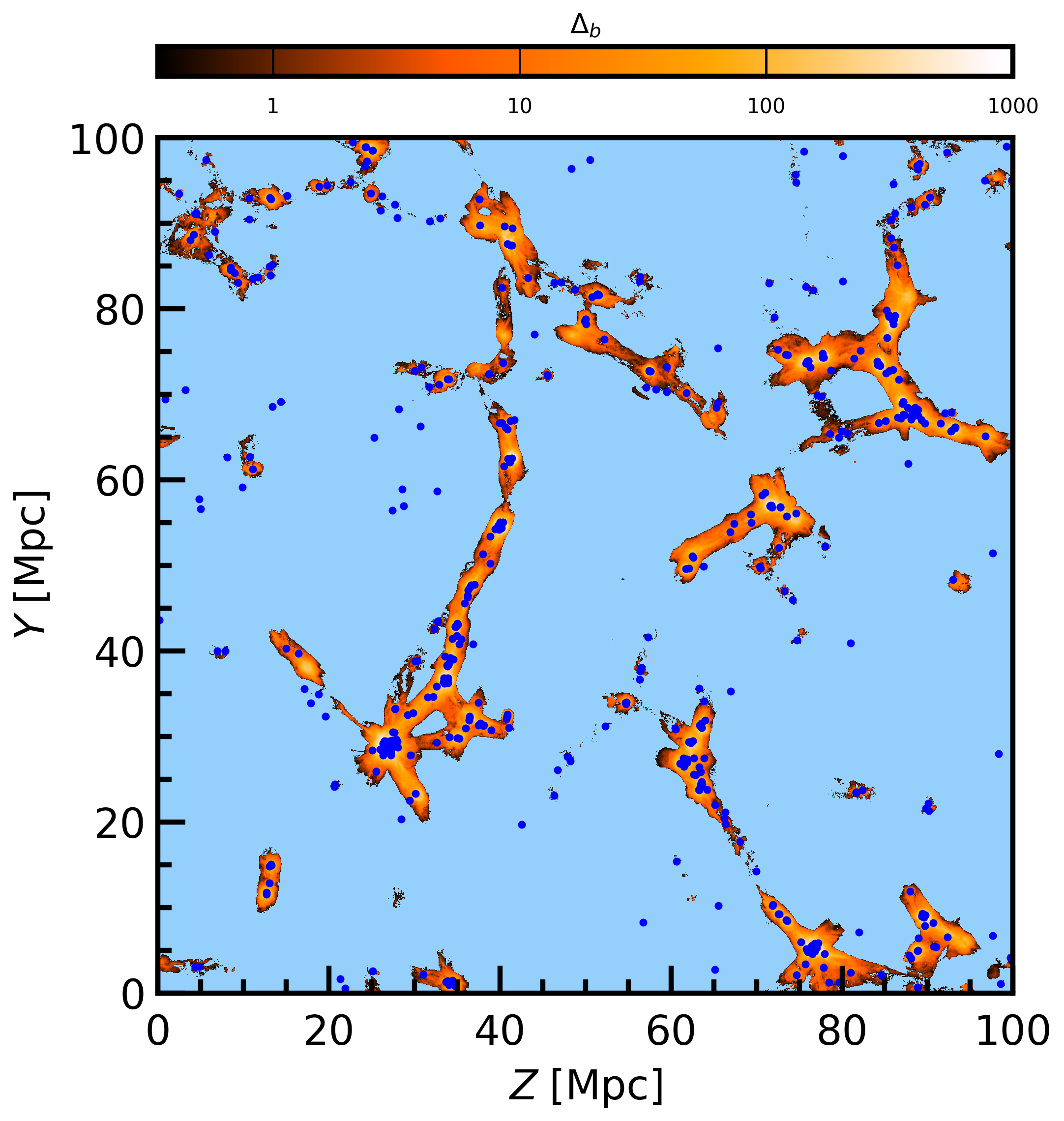}
\includegraphics[width=9cm,angle=0]{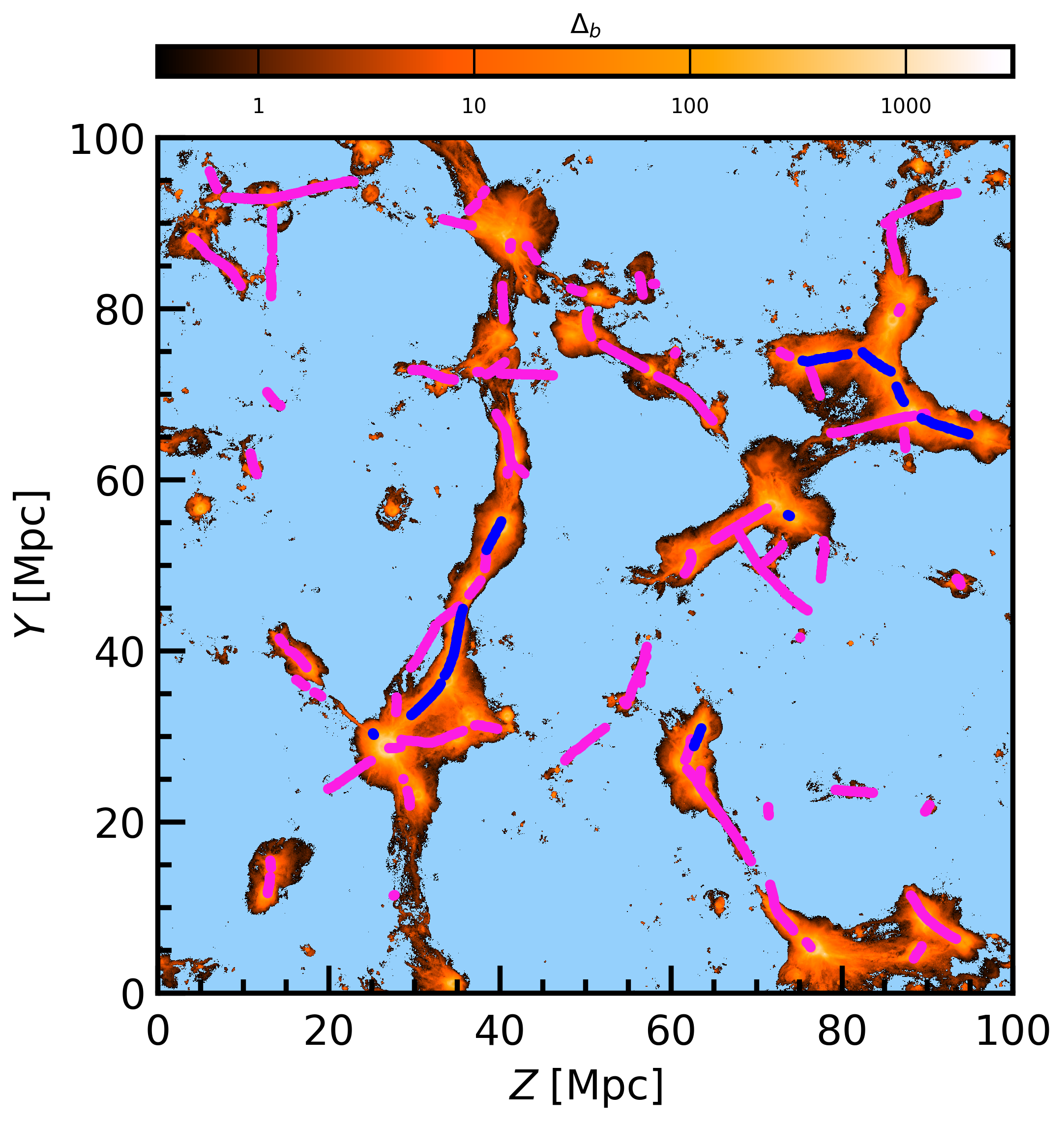}
}
}
\caption{Same slice as in Fig. \ref{full_slice_hr.fig} for visualisation of the Bisous formalism.
{\it Top left}: The galaxies (blue dots) and the visit map $\ge$ 0.05 (grey). 
{\it Top right}: The spatial distribution of the missing baryons (colour map) and the filament volumes (grey areas). 
{\it Bottom left}: The spatial distribution of the missing baryons captured with the Bisous filament finder.
{\it Bottom right}:  The spatial distribution of the missing baryons (colour map) and the filament spines of our high luminosity overdensity $\delta_{LD}$ sample (blue) and
filaments excluded from the high $\delta_{LD}$ sample (magenta).
}
\label{visit_gal_gas.fig}
\end{figure*}

\section{Overall spatial distribution of thermodynamic properties}
\label{spatio-thermal}
In this section we utilise the information on the spatial distribution of the baryons to improve the search for the missing ones.

\subsection{Temperature map}
\label{Tmaps}

Using the same representative 5 Mpc thick slice as in in Fig. \ref{full_slice_hr.fig}, we computed the mass-weighted mean temperatures of each particle in a projected 2D grid with a resolution of 0.1 Mpc. In order to visualise the temperatures of only the intergalactic medium, we excluded all particles within the virial radius R$_{200}$ of haloes. 
The resulting temperature map (Fig. \ref{tmap.fig}) reveals a significant structure. Most of the volume appears to have low temperatures ($\log{T(K)} \le 5$), presumably corresponding to the voids and sheets.

  Importantly for the current work, at the WHIM temperatures ($\log{T(K)} = 5-7$) the Cosmic Web - like structure appears. This is expected from previous simulations \citep[e.g.][]{2001ApJ...552..473D, 2006MNRAS.370..656D,2009ApJ...697..328B,2012MNRAS.425.1640T,2012MNRAS.423.2279C, 2018MNRAS.473...68C, 2019MNRAS.486.3766M} and supports our approach of searching for the hottest WHIM within the cosmic filaments. 
  The temperature structures have very sharp edges, revealing the accretion shock heating processes which are essential in our adopted theoretical scenario.
  
  At the highest temperatures ($\log{T(K)} \ge 7$) the mass is concentrated around the crossing points of the apparent filaments traced by the $\log{T(K)} = 5-7$ gas.
  They are the sites of the most energetic gravitational collapse induced shock heating.

\subsection{Densities of different phases}
\label{phase_slices}
We repeated the procedure of using the representative slice of the simulation box from Section~\ref{EAGLEbar}, but this time constraining the baryon temperatures of gas outside R$_{200}$ into different ranges.
Visual inspection (see Fig.~\ref{phase_slices.fig}) yields the following suggestions. We will investigate the suggested behaviour of the temperature distribution in different Cosmic Web environments quantitatively with NEXUS+ as a function of temperature in Section~\ref{nexus}. In order to roughly approximate the volume filling fractions, we computed the mass-weighted mean temperatures of the particles in the representative slice (see Fig. \ref{full_slice_hr.fig}) in a 3D grid, with a separation of 0.1 Mpc between grid points. The volume filling fraction for each environment was then calculated by dividing the number of grid points associated with the respective temperature range by the total number of grid points in the representative slice.

\begin{itemize}

\item
  The baryons with $\log{T(K)}~<~3$ trace the lowest-density parts of the voids. The absence of such cold gas between the voids suggests the Cosmic Web structure. This phase occupies the largest volume at $\sim$~55\% of the total volume. 

\item
  At $\log{T(K)}~=~3-4$ the volume filling factor is smaller, $\sim$~23\%, consistent with the assumption that this temperature range presents the hottest (i.e. rare) void gas together with sheets which are less volume filling but warmer than the voids. 

\item
  The baryons with $\log{T(K)}~=~4.0-5.0$ correspond to the bulk of the intergalactic phase detected via the Lyman alpha forest. The volume filling fraction ($\sim$~12\%) is smaller than at the lower temperatures since the neutral hydrogen, accumulated towards the more compact filaments and sheets, becomes important.

\item
  In the warm FUV-detected range of $\log{T(K)}~=~5.0-5.5$ the mass is much more concentrated, as demonstrated by the smaller volume filling fraction of $\sim$~5\%.  While the sheets still contribute to this temperature range, the mass is now more concentrated around the apparent galaxy filaments which are warmer than the sheets due to the low energy part of the filament accretion shock-heated gas. 

\item
  In the missing baryon range (i.e. $\log{T(K)}~>~5.5$ excluding gas within R$_{200}$) the diffuse gas is compact with a volume filling fraction of $\sim$~5\% and very strongly concentrated to the apparent filaments.
    This is consistent with our adopted scenario whereby this phase is produced by the strongest accretion shocks due to mass flow towards filaments.

\item
  The virial structures as defined in Section~\ref{bound} (dominating the mass budget at the highest temperatures of $\log{T(K)}~>~6$) apparently trace the same general large-scale web structure as the WHIM, but have a very narrow
and clumpy spatial distribution within the apparent filaments. This phase is produced by the gravitational collapse and mergers of dark matter haloes.

\end{itemize}

\begin{figure}
\includegraphics[width=9cm,angle=0]{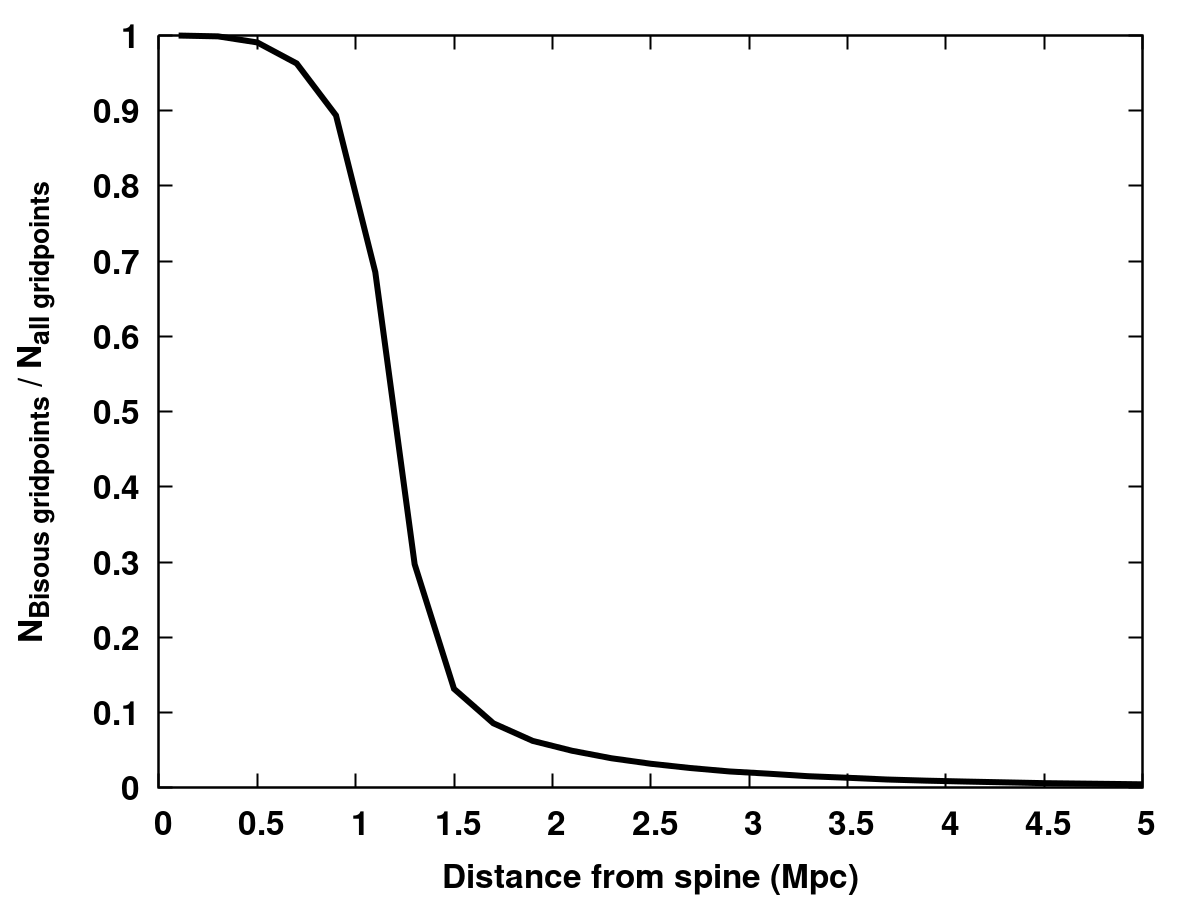}
\caption{Fraction of grid points within Bisous volumes (visit map value > 0.05) as a function of distance from the filament spines. All spines were stacked together.}
\label{bisous_radius.fig}
\end{figure}

\section{The filaments}
\label{fila}
\subsection{The detection method (Bisous)}
\label{bisous_detection}
Our goal is to investigate the properties of the missing baryons in the environment of the cosmic filaments.
To enable this, we applied the Bisous formalism to EAGLE galaxies to detect the filaments (see \citealt{2007JRSSC..56....1S,2010A&A...510A..38S,2016A&C....16...17T} for a detailed description of the method).
It is designed for detecting the large-scale filamentary networks based on spectroscopic galaxy catalogues (both observational and simulated).
We used the software to model the distribution of the EAGLE galaxies brighter than M$_r$=-18.4 (see Section~\ref{gal}) by fitting connected and aligned cylinders to it. We repeated this 1000 times to account for the stochastic
nature of the Bisous formalism. We thus obtained a map for each run, indicating the spatial volumes covered by the fitted cylinders (hereafter we will refer to this map of filament coverage as a visit map).
The regions which get covered more often in the individual fits are more likely to belong to a filament. Our experimental definition of a spatial location belonging to a filament is that it is covered by more than 5\% of the individual visit maps, since this choice adequately yielded the statistical properties of the SDSS filaments \citep{2014MNRAS.438.3465T}.
We define the filament spines as the local maxima of the combined visit map, i.e. where most of the separate visit maps overlap.

In this work, we employed the same version of the Bisous as used for detecting the filamentary galaxy network in SDSS \citep{2014MNRAS.438.3465T}
(see the above references for details of the method).
Here, however, the cylinder radius is not fixed to 0.7 Mpc, as in  \citet{2014MNRAS.438.3465T},  but is selected to be in the range 0.5 - 1.0 Mpc, as in \citet{2020A&A...639A..71K}.
Thus, the Bisous approach adopted in this paper is not optimised for detecting gas or the missing baryons, but to construct the filamentary network from the distribution of galaxies. In fact, a recent study by \citet{2019MNRAS.490.1415K} found that the aforementioned radius for the Bisous filaments was appropriate to obtain a reliable signal-to-noise ratio for neutral \ion{H}{I} observations. Our aim in this work is to evaluate the efficiency of the current version in tracing the hot WHIM and, if promising, we will develop it for observations in future work.

\begin{figure}
   \begin{minipage}[b]{0.45\textwidth}
    \includegraphics[width=7.5cm,angle=0]{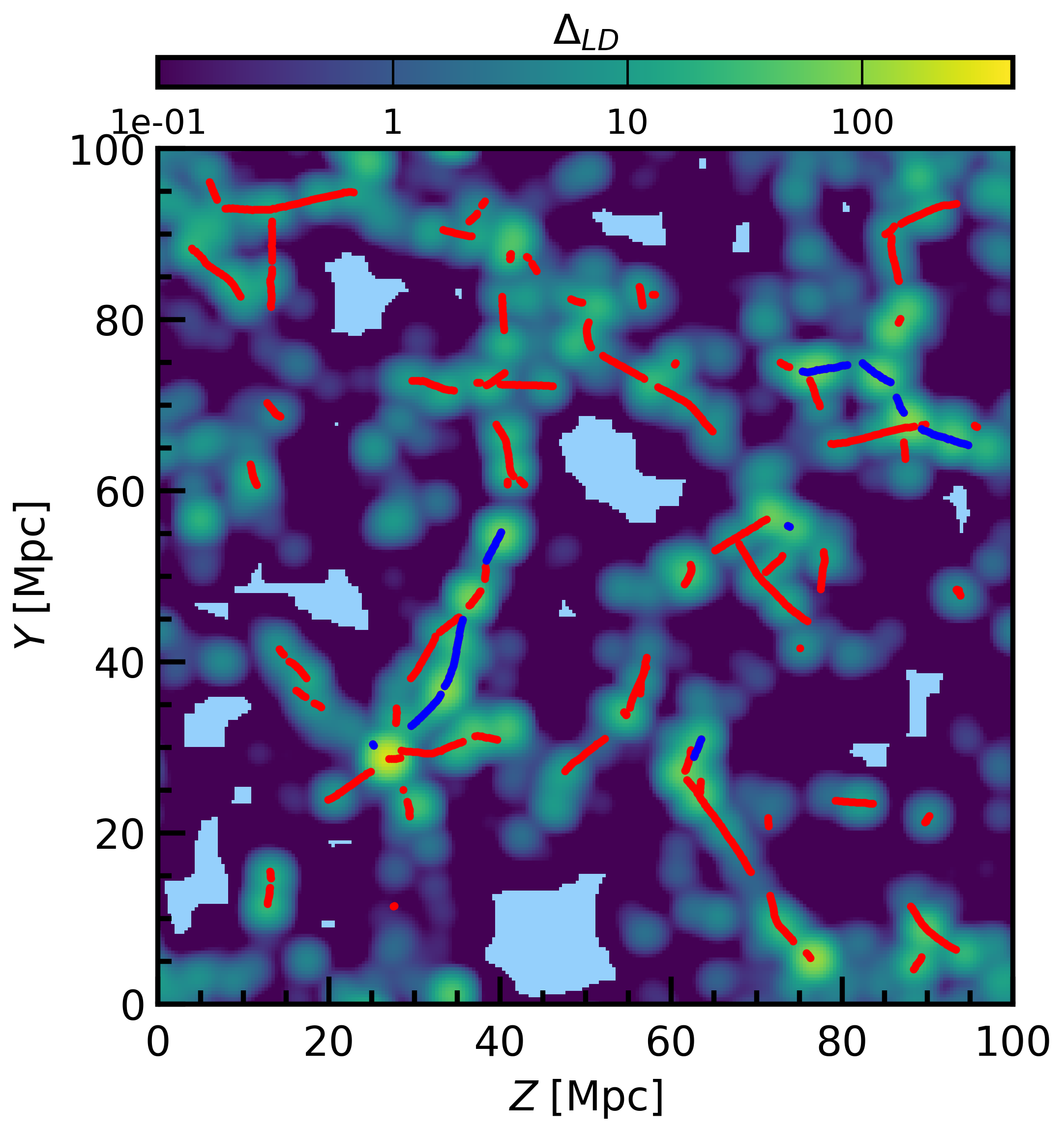}
    \caption{Galaxy luminosity density contrast ($\Delta_{LD}$) in the same slice as in Fig. \ref{full_slice_hr.fig}, using a 2 Mpc smoothing kernel. Filament spines of the high $\delta_{LD}$ sample are shown in blue, the excluded filaments are shown in red. }
    \label{LD_map.fig}
  \end{minipage}
   \hfill
   \begin{minipage}[b]{0.45\textwidth}
     
     \includegraphics[width=9cm,angle=0]{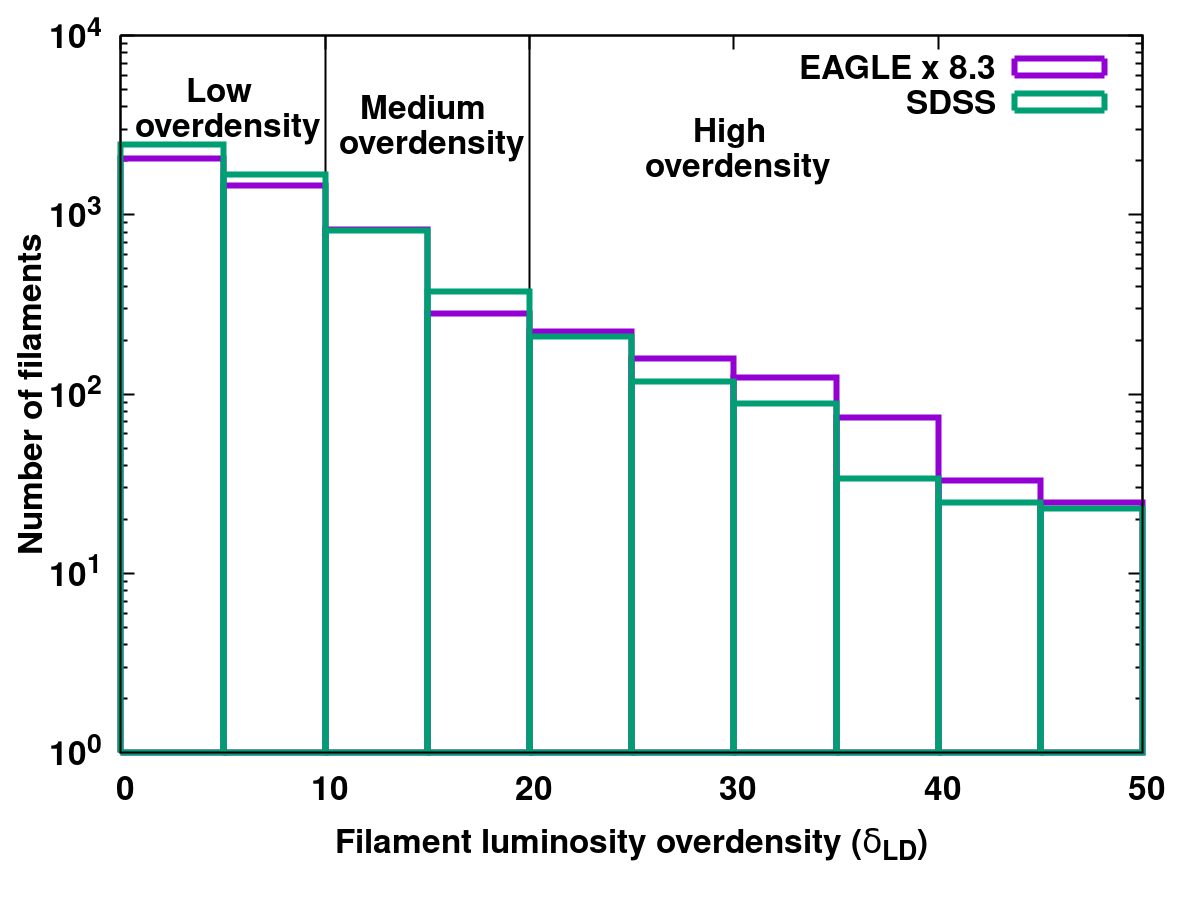}
     \caption{Distribution of filaments as a function of the luminosity overdensity in SDSS (green line) and in EAGLE (purple line, normalised to SDSS volume).
 The vertical lines indicate the limits for the low ($\delta_{LD} = 0-10$), medium ($\delta_{LD} = 10-20$) and high ($\delta_{LD} = 20-50$) luminosity overdensity groups (see Section \ref{classification}).}
    \label{LD.fig}
   \end{minipage}
  
\end{figure}

\subsection{Results}
\label{filres}
The visit map values (see Section \ref{bisous_detection}) were stored in a 3D grid with a separation between grid points of 0.4 Mpc \footnote{ We analysed how changes in the grid size would affect the capturing of missing baryons (see Section \ref{capt}). A grid size of 0.4 Mpc is optimal as it is not computationally demanding and the results have converged.}. The filament volume covered by the visit map $\ge$ 0.05 occupies $\approx$ 5\% of the EAGLE volume (see Fig.~\ref{visit_gal_gas.fig}), hosting $\approx$ 53\% of the baryon gas (including gas within collapsed structures). Excluding baryons within R$_{200}$, the Bisous filaments contain $\approx$ 40\% of the total baryon budget.
The filament spines were stored as a catalogue of spine points with their respective xyz-coordinates. In order to avoid possible edge effects, we excluded all the filament spines for which any spine point is closer than 3 Mpc to the edges of the simulated volume. The resulting sample contains a total of 779 filament spines (see Fig.~\ref{visit_gal_gas.fig}).

The current Bisous set-up is tuned for detecting filaments at an approximate scale of radius r $\approx 1$ Mpc \citep{2014MNRAS.438.3465T}. We used the data in hand to check the accuracy of the filament scale. Visual inspection of the Bisous volumes (see Fig. \ref{visit_gal_gas.fig}) indicates quite a complex geometry. Filaments cross and curve three dimensionally with both narrow and wide regions. Thus, to investigate the filament scale issue, we computed the stacked ratio of the grid points determined to be within a filament to the total number of grid points as a function of distance from the filament spines (see Fig. \ref{bisous_radius.fig}). We found that at r $ = 1$ Mpc distance from the filament spine on average $\approx$ 70\% of the volume probed by the grid points is within filaments. Subsequently the distribution experiences a steep drop, and at r $ = 1.5$ Mpc only $\approx$ 10\% of the grid points remain within the Bisous filaments. This quantitative result verifies that the current Bisous set-up is correctly fine-tuned for detecting filaments at scales of r $\sim1$ Mpc.

\subsection{Classification}
\label{classification}
The filament length is one of the few properties which can be derived relatively easily from the spectroscopic galaxy surveys.
However, the different methods for detecting the filaments yield systematic differences in the basic filament properties \citep[see e.g.][]{2018MNRAS.473.1195L,2020A&A...642A..19M}. Also, by varying the parameters of a given filament detection method, one may obtain systematic differences in the extracted populations. Thus, the length of a filament spine is generally not a robust way of classifying the filaments.

Particularly, the Bisous formalism we are employing in this work with the parameter set optimised for SDSS may break a given structure, defined as a single filament by the DisPerSE method \citep{2011MNRAS.414..350S}, into several shorter segments.
Also, the EAGLE simulation box size of 100 Mpc will suppress structure formation modes larger than this.
Consequently, the maximum Bisous filament spine length in this work is $\sim$~35 Mpc, shorter than in \citet{2020A&A...642A..19M}.

The gas density is a better filament measure, especially for studying the WHIM, as it is assumed to follow the local DM density. While the gas density is not directly available from the basic filament detection analysis, we can use the easiest and most direct measurable
property of the cosmic filaments, the galaxy luminosity, as a proxy. Namely, \citet{2015A&A...583A.142N} proposed that the luminosity density (LD) field of the galaxy population related to a given filament can be used as a low-scatter proxy for the WHIM density. The LD field method consists of three-dimensional smoothing of the spatial distribution of the luminosities of a given galaxy sample and evaluating the field at desired locations (see \citealt{2012A&A...539A..80L} for details of the method).

Indeed, to first order, we expect galaxies to trace the gas in filaments because they form in haloes collapsed into the filaments due to the same gravitational force as the more diffuse IGM. At the same time, the galaxy luminosities trace the masses of their parent haloes, and haloes of different masses form in regions with different large-scale over- and underdensities. Therefore, we utilised this scenario by employing the LD value as a filament classification criterion.
We applied the LD field method to the r-band luminosities of the EAGLE galaxies brighter than M$_r$ = -18.4 (see Section \ref{gal}), using a smoothing kernel width of 2~Mpc (see Fig. \ref{LD_map.fig}). 
We then evaluated the LD field at the locations along the spines of the 779 filaments detected in the EAGLE box. We divided the LD values by the average LD of the full EAGLE box ($\left<LD\right> = 7.1 \times10^{7}$ L$_{\odot}$ Mpc$^{-3}$) and consequently discuss the luminosity overdensity $\delta_{LD}~=~LD/ \left< LD \right> - 1$. For each filament we then calculated the $\delta_{LD}$ value by averaging the individual  $\delta_{LD}$ values of the spine points.

The resulting $\delta_{LD}$ distribution (see Fig.\ref{LD.fig}) exhibits a well-behaved decreasing trend towards high values up to 
$\delta_{LD}\sim50$. Above this, there is a low level tail extending to $\delta_{LD}\sim200$. Most of the highest values are due to filament spine fragments in galaxy groups. Our Bisous formalism is not optimal for working in such environments: some spines are spuriously created due to the local overdensity of galaxies without any filamentary structures being present. Thus, we experimented by limiting the further analysis to filaments with $\delta_{LD} \thinspace \le$~50. 

We then experimented by combining a modest length cut for the filament spines together with the luminosity overdensity $\delta_{LD}$. We found that a 2 Mpc cut is optimal: a deeper length cut reduces the number of selected filaments, while a more relaxed cut increases scatter due to spurious filament spines.

\begin{figure*}
  \vbox{
    \includegraphics[width=18cm,angle=0]{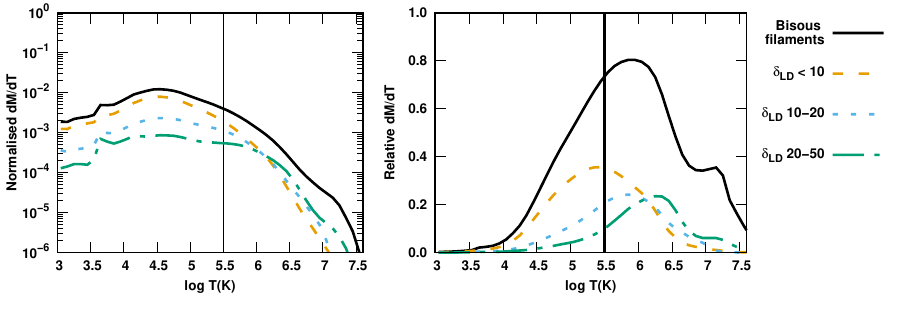}
    \includegraphics[width=18cm,angle=0]{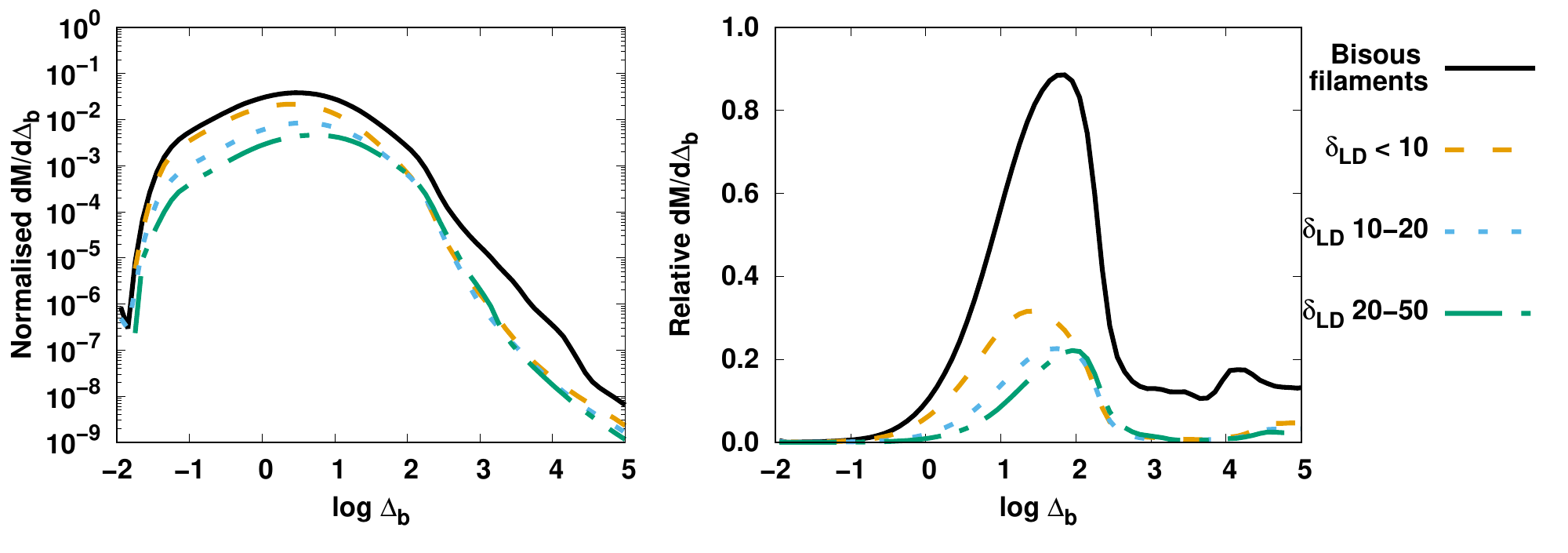}
}
  \caption{{\it Top:} Absolute (left panel, normalised to unity at the maximum value of the full baryon sample, left panel in Fig. \ref{mass_temperature.fig}) and relative distributions (right panel) of the baryon gas mass distribution as a function of mass-weighted temperature in filaments with low (orange dashed line), medium (blue dotted line) and high (green dash-dotted line) luminosity overdensity. For comparison, mass within all Bisous filaments is shown with a black line.
    {\it Bottom:} Same as top panel, but for the gas mass distribution as a function of gas density contrast. Gas within R$_{200}$ has been excluded in all panels.
}
\label{mass_temperature_LD.fig}
\end{figure*}

After excluding all the filament spines shorter than 2 Mpc, we divided the remaining spines into three groups based on their $\delta_{LD}$: low ($\delta_{LD} < 10$), medium ($\delta_{LD} = 10-20$) and high ($\delta_{LD} = 20-50$). The low, medium and high overdensity groups contain 432, 134 and 77 filament spines, respectively.
The relative overdensity distribution shifts towards higher values with higher luminosity density (see Fig. \ref{mass_temperature_LD.fig}), consistent with the relation between filament gas density and luminosity density, as described by \citet{2015A&A...583A.142N}. Since the observational signal improves with higher density, the above indicates that it is beneficial to concentrate the observational efforts on the high $\delta_{LD}$ filaments.

Bearing in mind that we want to develop methods to be applicable to observational data, we experimented whether the above LD classification scheme would work for the filaments detected in SDSS \citep{2014MNRAS.438.3465T}. We selected the same sample of filaments as published in \citet{ 2020A&A...639A..71K}, except for the detail that we excluded filaments shorter than 2 Mpc, not 5 Mpc (see the above work for the details of the analysis). 
Adjusting linearly with the difference of the volumes covered by SDSS in the adopted redshift range of 0.02-0.05 and EAGLE, we found that the $\delta_{LD}$ distributions for SDSS and EAGLE are very similar (see Fig. \ref{LD.fig}).
At the lowest luminosity densities the SDSS filament numbers agree with the scaled EAGLE values within 15\%. While both samples display a decreasing trend with higher luminosity density, SDSS drops a bit faster, resulting in a $\sim20$\% deficit
of the filaments in the high LD group. The larger difference in the luminosity distributions between EAGLE and SDSS galaxies (see Fig. \ref{teet.fig}) is not seen in the filament distribution, since most of the galaxies have magnitudes in the range where the two luminosity distributions agree. Thus, we assume that the deficit in SDSS filaments is small enough that the conclusions derived for the high $\delta_{LD}$ group in this paper are relevant for SDSS and that 
our filament classification scheme is applicable to SDSS at low redshifts. From a total of 5992 filament spines, the number of the SDSS filaments in the three $\delta_{LD}$ groups are 4247 for the low, 1190 for the medium and 498 for the high overdensity. We will perform a stacking analysis of the Planck data related to these filament groups in a future work. 

\section{Capturing the missing baryons with the filament finder}
\label{capt}

\subsection{Capture fraction}
\label{capture}

We extracted the total EAGLE baryon population within the full filament sample (see Section~\ref{fila}) for closer inspection. 
Applying our criteria for the missing baryons ($\log{T(K)} > 5.5$ and outside R$_{200}$, see Section \ref{definition}) to the above sample we then selected the missing baryon population captured by the Bisous filament finder. We thus define the ratio of the total mass of this sample to that of the total missing baryons in the full EAGLE volume (P$_{true}$, see Section \ref{definition}) as the capture fraction, i.e. the content of missing baryons that reside in the intergalactic space within filaments. According to EAGLE and the Bisous analysis, this capture fraction is $\approx$ 87\%.
 
Since the filaments occupy only $\approx$ 5\% of the full simulation volume (see Section~\ref{fila}), the above results on the EAGLE data demonstrate that the missing baryons are strongly concentrated towards the filaments.
This is consistent with the accretion shock heating scenario  \citep{2003ApJ...593..599R, 2005ApJ...620...21K}.
Since the Bisous formalism is applicable to observational galaxy surveys, our tests indicate that by focusing on the galaxy filaments detected in optical data one can localise most of the missing baryons.

The fraction of the captured missing baryons is comfortably high, especially considering that our current implementation of the Bisous formalism 
is optimised for the SDSS galactic filament detection and not for gas detection (see Section~\ref{fila}).
In particular, we are missing some of the more extended hot gas in the outskirts of the thickest filaments which is less strongly correlated with the galaxies (see Fig.~\ref{visit_gal_gas.fig}).

\subsection{Thermodynamic properties}
\label{thermodyn}
To investigate the thermodynamic properties of the baryon population within Bisous filaments, we analysed the corresponding mass distribution dM/dT and phase diagram (similarly to the collapsed phase, see Section \ref{bound}). While the collapsed phase dominates the higher temperatures, the gas captured by the Bisous filaments is most abundant at temperatures $\log{T(K)} = 5-6.5$ (Fig. \ref{mass_temperature_LD.fig}). Within this temperature range, $\approx$ 80\% of the gas in filaments resides outside R$_{200}$. The distribution peaks within the missing baryon phase (described in Section \ref{definition}), at a temperature of $\log{T(K)} \sim 5.9$. Above the missing baryon temperature limit ($\log{T(K)} = 5.5$) and outside R$_{200}$, the Bisous filaments contain $\approx$ 25\% of the total baryon budget (see Table \ref{eagle_budget.tab}).  Since this diffuse gas is hard to observe, it fits well in the role of the missing baryons. Its detection would constitute a crucial step towards the solution of the cosmological missing baryons problem. 

The temperature-density phase diagram of the Bisous filaments (Fig. \ref{phasediag.fig}) indicates that while the density contrast varies greatly from the lower end ($\log{\Delta_{b}} \sim 0$) to the higher end ($\log{\Delta_{b}} \sim 4$), most of the gas is concentrated within $ \Delta_{b} \sim 10-150$.
This density concentration corresponds to the temperature concentration at $\log{T(K)} = 5-6.5$ reported above, rendering this phase quite compact and smooth in the temperature-density plane.

The hottest and densest phase within Bisous filaments resides within the outskirts of galaxies and groups of galaxies (see Section \ref{bound}). 
The lower density and temperature phase of the Bisous baryons forms a progressive transition from the high density and temperature virial and outskirt phases  (see Fig.~\ref{phasediag.fig}), albeit with some overlap between the different phases.

The baryon sample extracted within the filaments also contains already robustly detected lower temperature phases 
(see Fig.~\ref{mass_temperature_LD.fig}). The ratio of the missing baryon mass to the total baryon mass within the filaments, i.e. for our purposes the purity, in the full EAGLE filament sample is $\approx$ 63\%. Our experiments with the EAGLE data and Bisous filaments indicated that by excluding the shorter filaments we do not improve the purity significantly while the capture fraction decreases.
This is due to the lack of correlation between the gas temperature and the Bisous filament spine length. This indicates that the filament length (which can be derived relatively easily from the galaxy spectroscopic measurements via the Bisous filament finder, see more discussion in Section~\ref{classification}) alone is not useful for the selection of the most missing-baryon rich filaments.

Thus, we analysed the thermodynamic properties of the $\delta_{LD}$ groups described in Section \ref{classification}. We found that the purity increases with $\delta_{LD}$, reaching a level of $\approx$ 82\% for the high density class.
This is due to the increasing relative contribution from the hotter gas with higher $\delta_{LD}$  (see Fig. \ref{mass_temperature_LD.fig}). This re-enforces our previous statement about the need to focus on the filaments with a high luminosity density when searching for the missing baryons (Section \ref{classification}).

\begin{figure*}
  \vbox{
    \includegraphics[width=18cm,angle=0]{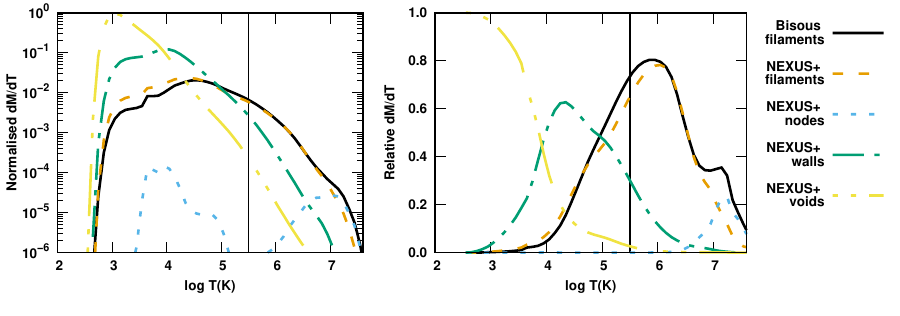}
    \includegraphics[width=18cm,angle=0]{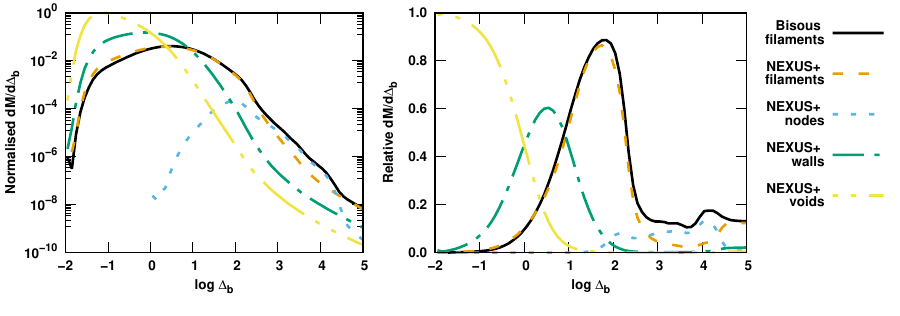}
    }
  \caption{
    {\it Top:} Absolute (left panel, normalised to unity at the maximum value of the full baryon gas sample, same as figures \ref{mass_temperature.fig} and \ref{mass_temperature_LD.fig}) and relative distributions (right panel) of the gas mass as a function of the mass-weighted temperature for Bisous filaments (black continuous line), NEXUS+ filaments (orange, dashed line), NEXUS+ nodes (blue, dotted line), NEXUS+ walls (green, dash-dotted line) and NEXUS+ voids (yellow, dot-dashed-dotted line).
    {\it Bottom:} Same as the top panel, but for the density contrast. The gas within R$_{200}$ has been excluded in all panels. }
\label{NEXUS_BISOUS.fig}
\end{figure*}


\subsection{Extending the view with NEXUS+}
\label{nexus}
When deriving observational implications for the diffuse hot cosmic baryons with our approach, we should estimate the systematic uncertainties related to our essential assumptions that
1) the Bisous filament finder method we employed does accurately capture the hot baryon population related to the cosmic filaments
and 
2) the data in the EAGLE simulation we utilised does form an accurate representation of the hot baryons in the local Universe.
In order to isolate the two effects one should perform filament detection with several methods (including Bisous) to several simulations (including EAGLE) and investigate the hot baryon content extracted from the different filament samples. 
Such a substantial amount of work is beyond the scope of this paper.

To get an idea of the level of the systematics in our results due to the filament detection methodology, we performed one test on the same EAGLE data we have been using throughout this paper, but applying a very different method,
NEXUS+ \\
\citep{2013MNRAS.429.1286C} instead of Bisous, to dark matter density fields instead of the galaxies.
NEXUS+ uses the morphology of the density field deformation tensor (Hessian) as a basis for assigning each location of the large scale environment to a void, wall, filament or node.
On the other hand, the Bisous method uses the spatial distribution of galaxies to detect the filamentary network. An important difference between the two methods is that NEXUS+  is a scale-free algorithm, whereas Bisous requires
the specification of scales on which to search for the filaments. 

The definition of the boundary of a filament is essential when capturing the hot diffuse baryons embedded in the filaments, and it is very different in NEXUS+ and Bisous.
The main setup of NEXUS+ includes the following steps. First, a log-Gaussian filter is applied to a given density field at several smoothing scales. Second, the Hessian matrix of this field is calculated at each smoothing scale and the eigenvalues $\lambda_{1} \le \lambda_{2} \le \lambda_{3}$ of this Hessian are obtained. 
In the third step each point is assigned a morphology based on the inequalities of the eigenvalues. Nodes are defined as regions with $\lambda_{1} \approx \lambda_{2} \approx \lambda_{3} < 0$, filaments as regions with similar densities along two axes, $ \lambda_1 \approx \lambda_2 < 0 $ with $ \lambda_2 \ll \lambda_3$  and walls are regions with $ \lambda_1 \ll \lambda_2 $ and $\lambda_1 < 0$. Any volume element that cannot be characterised as a filament, wall or node is classified as a void. 
This process is repeated for every smoothing scale and in the final step, the information is combined to obtain a scale-independent signature. Morphological features with a signature above this threshold are considered as valid structures. This scale-space analysis ensures that the morphological features prevalent on several spatial scales are identified. 
As a result, NEXUS+ gives a volume-filling field with each point labelled by the local morphology: node, filament, wall and void.

On the other hand, in our current implementation of the Bisous method the location of the filament boundary essentially depends on our choice of a spatial location being assigned to a filament if it is covered
by more than 5\% of the visit maps indicating the locations covered by the fitted cylinders (see the full description in Section \ref{fila}).
If we defined the filament volume by requiring a higher number of Bisous cylinders to cover it, we would shrink the filament boundaries closer to the filament axis and thus capture smaller fraction of missing baryons.
We will revisit this issue when deriving conclusions below.

While the two particular methods we are comparing here work with distinct mass components (NEXUS+ traces dark matter and gas; Bisous works on galaxies), they are expected to be spatially well correlated.
Namely, the dark matter dominates the formation of the large-scale structure and drives the formation of galaxies embedded within the Cosmic Web  \citep[e.g.][]{2013MNRAS.429.1286C}.
Also, the diffuse baryons are expected to be concentrated around the dark matter density maxima of the large scale structure, i.e. the axes of the Cosmic Web filaments. 
In fact, the volumes covered and the amount of dark matter, gas and stellar mass contained by the three different filament samples were consistent within 10\%.
These agreements suggest that 1) the NEXUS+ and Bisous methods quite accurately detect the main filamentary network and 2) dark matter, gas and galaxies quite accurately trace the same filaments.

Our specific focus in the current paper is on the $hot$ phase of the intergalactic gas in the filaments. Therefore, the above comparison of the full gas population is not adequate for quantitative assessment in our case. 
We thus utilised the above published characterisation of the large scale environments obtained by applying the NEXUS+ method
on the dark matter density fields in the EAGLE simulations \citep{2019MNRAS.487.1607G}.  
We repeated our analysis of the thermodynamic properties of the diffuse EAGLE baryons (excluding the gas within R$_{200}$, see Figs. \ref{mass_temperature.fig} and \ref{mass_temperature_LD.fig}), but this time using NEXUS+ voids, walls, filaments and nodes.
The resulting distributions of the mass as a function of temperature and overdensity indicate that through the voids-walls-filaments-nodes sequence the temperature and density increase systematically (see Fig. \ref{NEXUS_BISOUS.fig}).
This is expected in the scenario of gravitational collapse and shock heating, where sheets, filaments and nodes are collapsed in 1, 2 and 3 dimensions, respectively. Each collapsed dimension increases the density and causes more shocks, which in turn induces more heating and higher temperatures. Voids, on the other hand, experience no collapse and thus remain at the low end of the distributions.

We found that the filaments detected by applying NEXUS+ to the dark matter density fields of EAGLE captured $\approx$ 79\% of the missing baryons (i.e. diffuse baryons with $\log{T(K)} > 5.5$, outside R$_{200}$).
This is somewhat smaller than the corresponding value obtained with Bisous ($\approx$ 87\%, see Section \ref{capture}).  At the lower temperature end of the missing baryon phase ($\log{T(K)} \sim 5.5-6$, see Fig. \ref{NEXUS_BISOUS.fig}), the difference can be attributed to walls. As discussed in \citet{2019MNRAS.487.1607G} $\approx$ 30\% of the Bisous filament galaxies are located in regions classified as walls by NEXUS+ in the EAGLE simulation. The temperature distribution of the EAGLE baryons (see Fig. \ref{NEXUS_BISOUS.fig}) shows that the NEXUS+ walls contain a significant amount of the hot baryons. The downside is that most of the gas in walls is cooler than $\log{T(K)} = 5.5$ and thus its inclusion unavoidably reduces the purity
i.e. the mass fraction of the hottest WHIM to all the IGM captured within the filaments.

At the high temperature end ($\log{T(K)} > 6.8$) the relative mass distribution within the Bisous filaments has a bump, due to the outskirts of galaxies and groups of galaxies (see Fig. \ref{NEXUS_BISOUS.fig}).
NEXUS+ filaments do not have this feature, probably because the regions we defined as outskirts (using the FoF and R$_{200}$ information, see Section \ref{bound}) are situated within NEXUS+ nodes, and thus the baryons in those regions are not captured within the NEXUS+ filaments.
This explains partly the lower missing baryon capture fraction compared to that obtained with the Bisous filaments.
The gas in the outskirts has very high purity which very closely compensates for the lower purity in the walls.

The mass distribution as a function of temperature in different NEXUS+ environments corroborates the visual inspection carried out in Section \ref{phase_slices} (see Fig. \ref{phase_slices.fig}). The lowest temperature range ($\log{T(K)}~<~3$) is dominated by voids. Between $\log{T(K)}~=~3-4$ voids are still the largest component, but with increasing contribution from walls. At $\log{T(K)}~=~4-5$ the walls dominate, while at $\log{T(K)}~>~5$ filaments emerge as the most prominent structures.

The density distributions of the intergalactic gas (outside R$_{200}$) in NEXUS+ and Bisous filaments are very similar (see Fig. \ref{NEXUS_BISOUS.fig}). As with the temperature distribution, on the high end of the density distribution the gas density within NEXUS+ filaments drops slightly faster than the density within Bisous filaments. This is explained by the gas contained in NEXUS+ nodes becoming more prominent towards higher densities.

In summary, assuming that the Bisous and NEXUS+ methods bracket the systematic effect related to the filament detection methodology, the total mass of the hot intergalactic baryons captured in the filaments may vary by $\approx$ 10\%. We consider this to be a good match, considering the very different approaches these two methods use to obtain filaments. While the purity remains almost the same, the Bisous method captures a higher amount of missing baryons resulting from the different galaxy populations belonging to the filaments traced by the two methods.
Nonetheless, both methods indicate that from all the hot gas in EAGLE ($\approx$ 42\% at $\log{T(K)} > 5.5$, see Table \ref{eagle_budget.tab}), $\approx$ 54\%-59\% reside in the IGM (outside R$_{200}$) within filaments.


\section{Profiles}
We next quantitatively analysed the visual regularities described in the previous section, whereby hotter and denser gas is closer to the filament spines, by deriving radial temperature and density profiles. For this end we used a simple, cylindrical geometry, where the gas properties were projected into a one dimensional distribution as a function of distance from the filament spine. For this aim we binned the filament gas within concentric hollow cylinders centred on each spine (for the radial binning, see Section~\ref{tprof}). We did not apply the filament volume definition (the visit map in Section \ref{fila}) here, as it would truncate the profiles at $\sim$1~Mpc (the number of particles per bin drops dramatically beyond this radius, see Section \ref{filres}). Instead, we were interested in seeing how the thermodynamic properties of the gas change when moving further away from the filaments.

We co-added the baryon masses and calculated the median baryon temperatures using the gas particles located within each hollow cylinder. We subsequently divided the masses by the hollow cylinder volumes to obtain densities in each distance bin. We repeated the above procedure for all filaments, thus obtaining a sample of 779 radial density and temperature profiles. We investigated the filament-to-filament variation of the density and temperature at each radius and include the scatter in our profiles (see Sections \ref{tprof} and \ref{dprof}).

We also performed a null test by repeating the profile extraction but this time changing the locations and orientations of the filament spines. We used the same 779 filament spines as before, but rotated each spine in a random direction and displaced it a random distance within the simulation volume. This way we eliminated any possible co-spatial connection between the filament spines and the gas distribution.

\subsection{Inner temperature profiles}
\label{tprof}
\begin{figure*}
\includegraphics[width=16cm,angle=0]{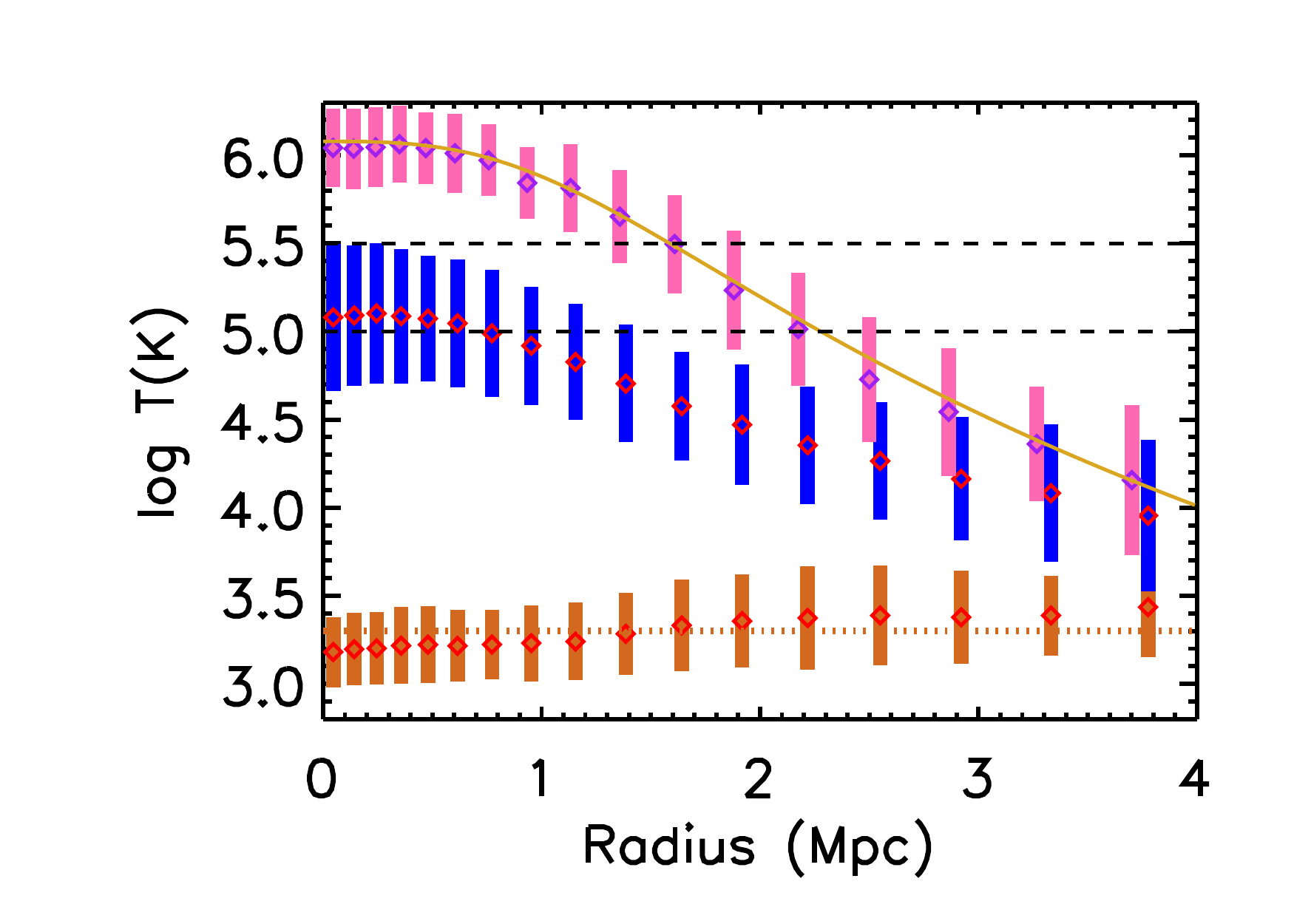}
\caption{Temperature profile as a function of distance from filament spines. The pink, blue and brown bars indicate the  1$\sigma$ interval of the log-normal distribution of the median values of individual filaments within each distance bin for the high $\delta_{LD}$, full and random filament samples, respectively. The red and purple squares mark the peak of each respective distribution. The solid yellow line is the best-fit temperature profile model fitted to the high $\delta_{LD}$ sample. The brown dashed line indicates the background temperature, while the black dashed lines mark the temperature limits of the WHIM ($\log{T(K)} = 5$) and the hot WHIM (i.e. missing baryons, $\log{T(K)} = 5.5$).}
\label{tprof.fig}
\end{figure*}

The resulting temperature distributions of one value per filament in each radial bin have a well-defined maximum and a symmetric distribution around it in logarithmic space.  
We thus fitted the distributions of the full sample with a log-normal model and used the best-fit parameters to construct the radial profiles of the most likely values and the 68\% confidence intervals of the temperatures (see Fig.~\ref{tprof.fig}). The resulting profile peaks at $r =0$ and decreases smoothly with increasing radius.

  The random profile sample (see above) indicated no significant radial temperature variation (see Fig. \ref{tprof.fig}) demonstrating that the central temperature increase in the filament sample is indeed associated with the filament spines.
The small increasing trend may be explained by the larger volumes at the outer radii more frequently capturing structures including hot gas. This also serves as an indication that the Bisous formalism we adopted for the filament detection is robust.   
  Since the median background value (T $ \approx 2  \times 10^{3}$ K) is below 10\% of the profile values at all radii, we ignored it in the profile fitting below.
  
The temperature profile derived from the full filament sample has a maximum of $\log{T(K)} \approx$ 5 at $r =0$. Thus, it is very unlikely to randomly encounter the hot WHIM temperatures.
We therefore repeated the above procedure using only the sub-sample of the filaments with the highest luminosity overdensity values ($\delta_{LD} =$ 20--50, defined in Section \ref{classification}).

In order to ensure proper counting statistics, we required that each hollow cylinder must contain more than 1000 gas particles.
Additionally, in order to ensure proper sampling, we required that the cylinder width be at least 5 times the mean interparticle distance $d$ given by the formula $d$ = (100/1504) / (1+$\delta_{b}$)$^{1/3}$ (in Mpc), where $\delta_{b}$ is the baryon overdensity computed for each particle. Considering both requirements, we increased the hollow cylinder width with increasing radius, from 0.1 Mpc in the centre to 0.4 Mpc at the largest radii.

As above, the maximum temperature occurs at $r =0$, but this time at a temperature of $\log{T(K)} \approx$ 6 (see Fig.~\ref{tprof.fig}).
This shows the importance of the high $\delta_{LD}$ sample: we probe the hot WHIM (i.e. the missing baryons) more efficiently, when focusing the observational efforts on the filaments with the highest luminosity overdensities.

Visual inspection of the temperature profiles (see Fig.~\ref{tprof.fig}) indicated a quite regular radial behaviour, justifying the employment of a radial profile model representing the high $\delta_{LD}$ sample. After experimenting with analytical functions, we settled on the commonly used single-$\beta$ 
model with a modification. We noticed that the model fits the data better if we change the exponent of the 
$\frac{r}{r_\mathrm{c,T}}$ term from 2 to 3, i.e. our fitting function was
\begin{equation}
T(r) = T_0 \times \left[1 + {\left( \frac{r}{r_\mathrm{c,T}} \right)}^3\right]^{(-\frac{3}{2} \beta_{T})} . 
\end{equation}
Since our profiles only extend to few times the core radius $r_\mathrm{c,T}$ (see below), we could not meaningfully constrain the slope parameter $\beta_{T}$. There is a high degree of degeneracy between $r_\mathrm{c,T}$ and $\beta_{T}$ and thus, after trial and error, we fixed $\beta_{T}$ to 1.0.
We then fitted T$_0$ and $r_\mathrm{c,T}$ to the data in the radial range $r$ = 0 -- 4 Mpc following this procedure: 
1) We used the best-fit log-normal models of the temperature distributions at each radius (obtained above) to derive random variables and thus constructed 1000 randomised temperature profiles\footnote{We found that the best fits had converged when using 1000 randomisations.}.
Such a sample statistically reflects the filament-to-filament variation.
2) We fitted each randomised temperature profile with the above model using a simulated annealing procedure \citep{press} to minimise the residuals between the model prediction and the simulated temperature value.
3) We fitted the distributions of the best-fit T$_0$ and
$r_\mathrm{c,T}$ values with a log-normal model (see Fig. \ref{tprof.fig}), thus yielding the most likely values and the 68\% confidence intervals.

The procedure yielded 
$T_{0} \thinspace = \thinspace 1.2_{-0.2}^{+0.3} \thinspace \times \thinspace 10^{6}$ \thinspace K , $r_{c,T} \thinspace = \thinspace 1.4_{-0.2}^{+0.3}  $ \thinspace Mpc (see Fig. \ref{tprof.fig}).
Our temperature profile is qualitatively similar to that of \citet{2019MNRAS.486..981G}, in that it has a rather flat central part and a steeper drop at larger radii. Quantitatively the match is not good: our values are lower at all radii and our core radius is smaller.

The temperature of the best-fit model decreases (Fig. \ref{tprof.fig}) below the approximate warm/hot WHIM border of $10^{5.5}$ K at $r~\approx$ 1.5~Mpc.
This demonstrates the importance of focusing the missing baryon search close to the filament spines.

\begin{figure*}
\includegraphics[width=16cm,angle=0]{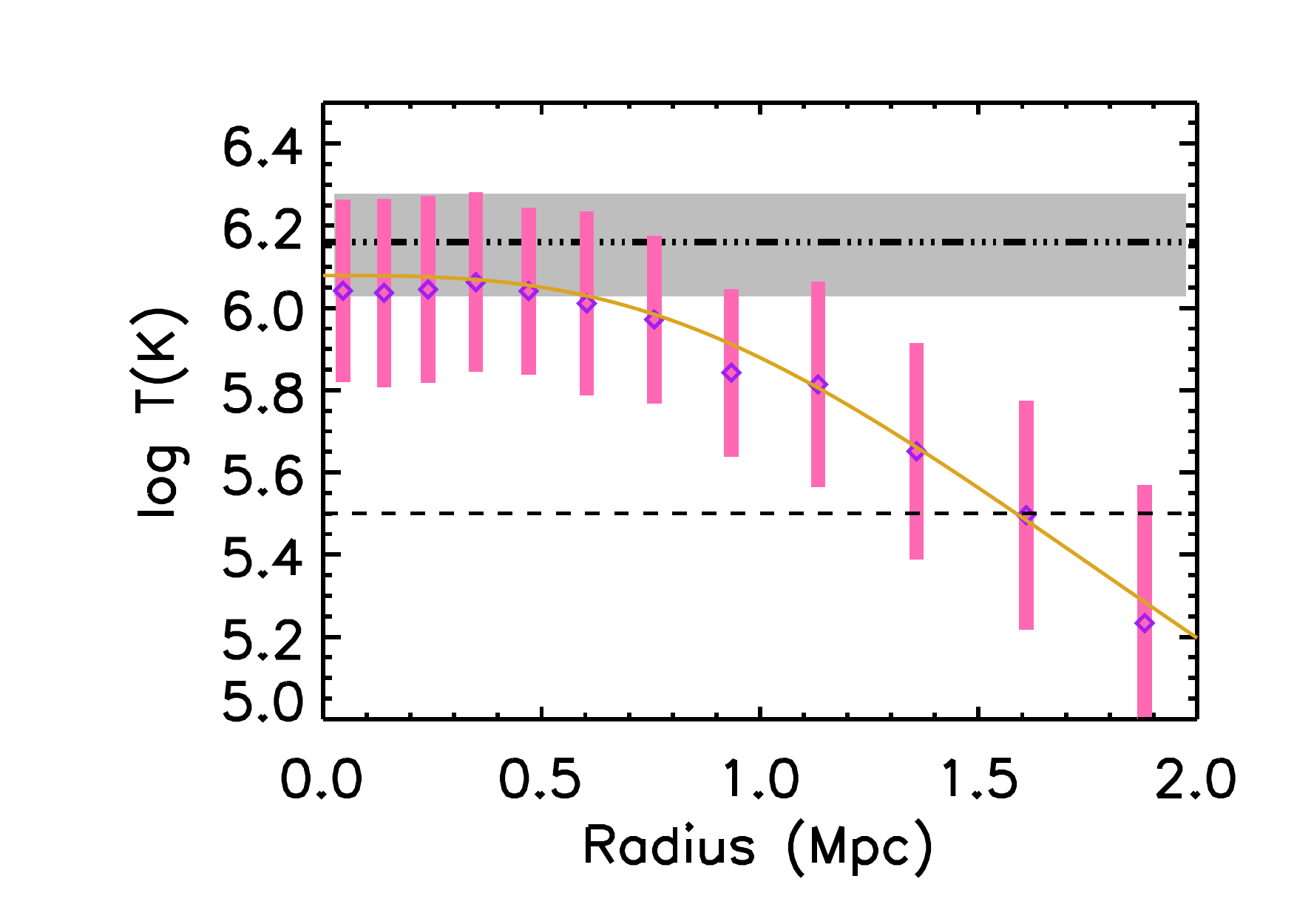}
\caption{Temperature as a function of distance from the filament spines. The solid yellow line is the best-fit profile for the high $\delta_{LD}$ sample with its respective distribution indicated by the pink bars (same as in Fig. \ref{tprof.fig}). The black dash-dotted line is the temperature derived by \citet{2020A&A...637A..41T} from Planck tSZ and CMB-lensing observations, with the grey area showing the error. The dashed line at $\log{T(K)} = 5.5$ indicates the lower limit for the hot WHIM.}
\label{tprof_comparison.fig}
\end{figure*}

We then used the observational results obtained by \citet{2020A&A...637A..41T} on the radial distribution of the filament gas temperatures for comparison with our results derived from the EAGLE simulation. Using Planck \citep{2016A&A...594A..22P} tSZ (thermal Sunyaev-Zeldovich) and CMB-lensing data, \citet{2020A&A...637A..41T} derived a constant temperature profile of $T \thinspace = \thinspace 1.4_{-0.4}^{+0.4}  \thinspace \times \thinspace 10^{6}$ \thinspace K up to $r  = 5$ Mpc. While \citet{2020A&A...637A..41T} used a different filament detection method (DisPerSE) for a different sample of SDSS filaments, their results are consistent with ours in the central $r \approx$ 1 Mpc region (see Fig. \ref{tprof_comparison.fig}). We further discuss the similarities and differences in Section \ref{disc}.

\begin{figure*}
\includegraphics[width=16cm,angle=0]{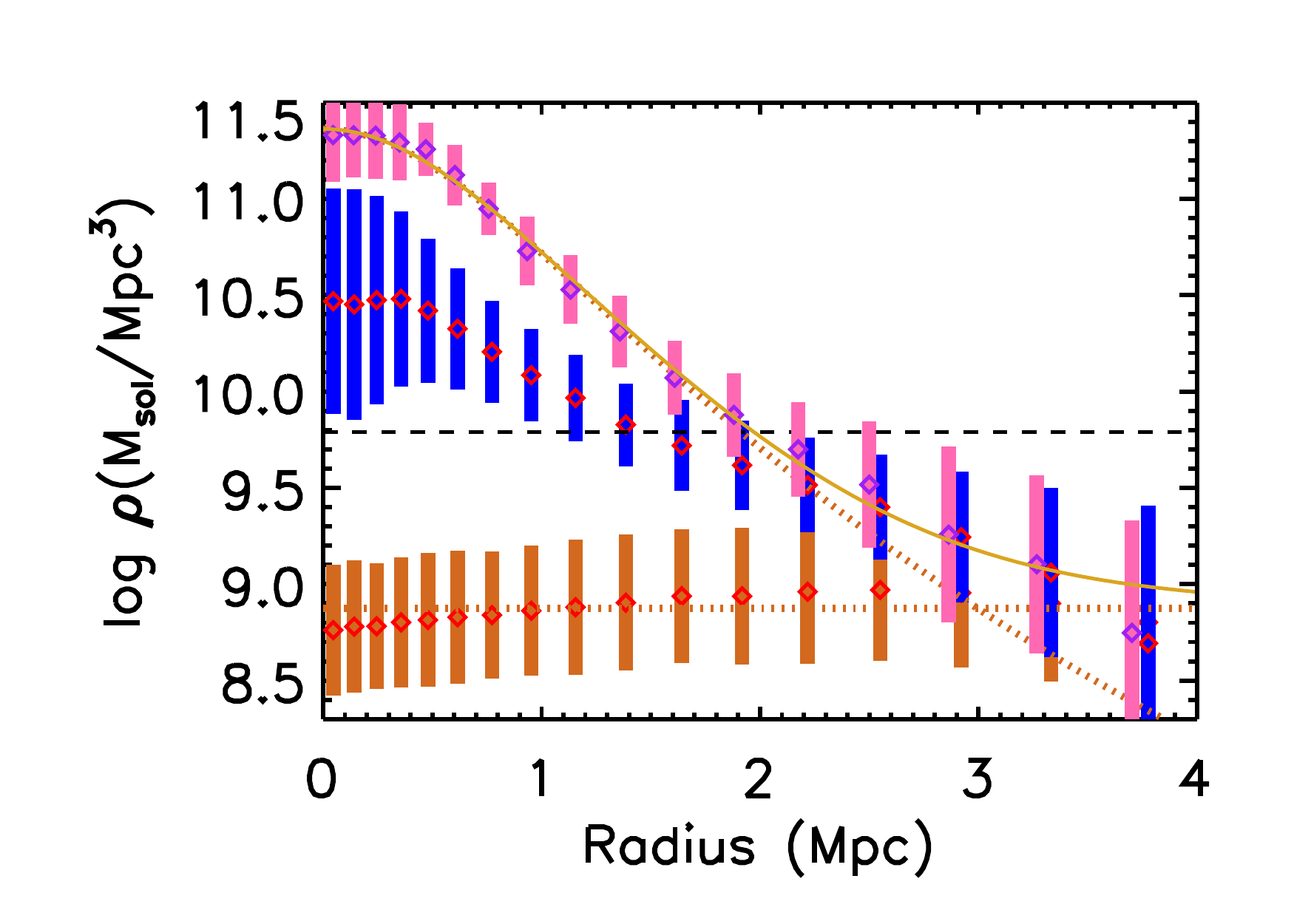}
\caption{Gas density profile as a function of distance from filament spines. The  1$\sigma$ interval of the log-normal distribution of individual filament densities at each distance bin for the high $\delta_{LD}$, full and random filament samples is represented by pink, blue and brown bars, respectively, with the squares indicating the peak of the distribution. The yellow line is the best-fit density profile model of the high $\delta_{LD}$ sample. The descending brown dotted line is the best-fit density profile model without the effect of the background. The black dashed line indicates the mean cosmic baryon density, while the brown dotted line shows the background level.}
\label{dprof.fig}
\end{figure*}

\begin{figure*}
\includegraphics[width=16cm,angle=0]{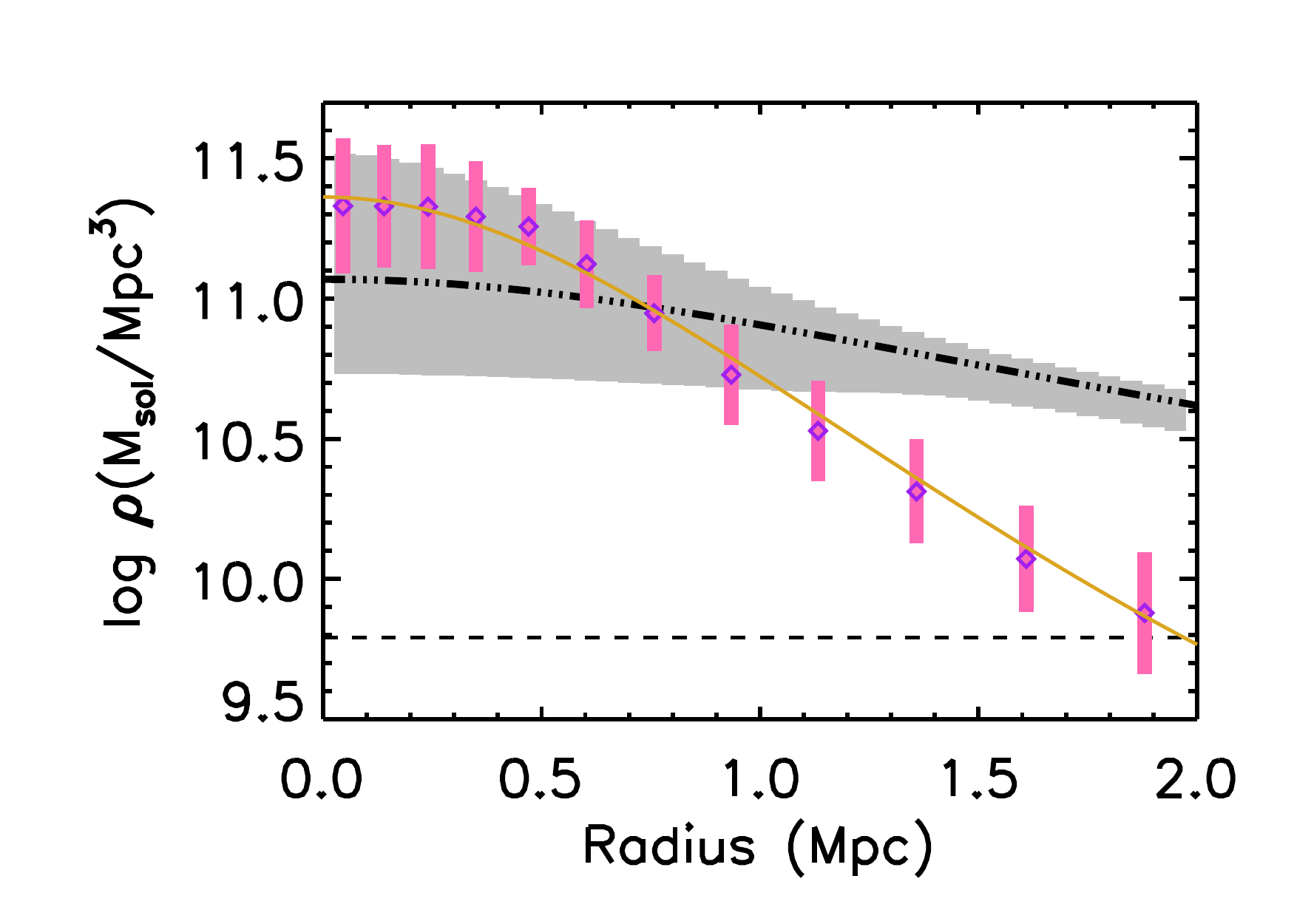}
\caption{Comparison between the baryon density profile of our high $\delta_{LD}$ sample (yellow line and pink bars) and the baryon density profile derived by \citet{2020A&A...637A..41T} (black dash-dotted line for the best-fit model and grey area for the error). In order to derive their density profile, \citet{2020A&A...637A..41T} fitted a beta-model profile to the tSZ and CMB-lensing data observed by Planck. The mean cosmic baryon density is represented by the black dashed line.}
\label{dprof_comparison.fig}
\end{figure*}

\subsection{Inner density profiles}
\label{dprof}
Similarly to the temperatures, the densities in each radial bin follow a log-normal distribution well. Following the above procedure, we determined the profiles of the most likely density and the 68\% confidence intervals for the full sample
(see Fig.~\ref{dprof.fig}).
The resulting profile has a maximum at $r =0$ and decreases smoothly with the increasing radius.

As in the case of the temperatures above, the null test (i.e. the random profile sample) indicates no significant radial density variation 
(see Fig. \ref{dprof.fig}), reinforcing the conclusions about the association of the physical WHIM properties with the filaments and the Bisous robustness. 
We adopt the mean value $\rho \approx 6.5  \times 10^{8}$ M$_{\odot}$/Mpc$^{3}$ as the background level for the profile fitting below.

The maximum baryon overdensity derived from the full filament sample is very low, $\delta_{b} \approx$ 5 at $r =0$. 
Thus, considering the full filament sample, it is very unlikely to randomly encounter the highest expected baryon overdensities in the 
cosmic filaments ($\sim$100).
We therefore repeated the above procedure using only the high $\delta_{LD}$, defined in Section \ref{classification}.
As above, the baryon overdensity is largest at $r =0$, but this time at a much higher value, $\delta_b \approx$ 40 (see Fig.~\ref{dprof.fig}). 
Thus, by focusing observational efforts on the filaments with the highest luminosity densities one probes the densest phase of the hot WHIM, 
i.e. the one with the highest X-ray and SZ signal.

Visual inspection of the density profiles indicated a quite regular radial behaviour, justifying the employment of a radial density profile model representing the high $\delta_{LD}$ sample.
We fitted the density profile using a single-$\beta$ model, usually used for the intracluster gas density:
\begin{equation}
\rho(r) = \rho_0 \times \left[1 + {\left( \frac{r}{r_{c, \thinspace \rho}} \right)}^2\right]^{(-\frac{3}{2} \beta_{\rho})} + bkg
\end{equation}
using the same randomisation scheme as above in the case of the temperature profiles.
Following the arguments given for the case of the temperature profiles, we fixed the slope parameter $\beta_{\rho}$, this time to 2.0.
Since the density of the filament sample approaches the background level at the largest radii, we included the constant background level (bkg) derived above
in our density profile model.

The procedure yielded $\rho_{0} \thinspace = \thinspace 2.3^{+0.6}_{-0.5} \thinspace \times  \thinspace 10^{11}$ \thinspace M$_{\odot}$ \thinspace Mpc$^{-3}$
and r$_{c, \thinspace \rho} \thinspace =  \thinspace 1.2^{+0.3}_{-0.2} \thinspace$ Mpc.
Our gas mass density profile is different from the gas particle density profile in the \citet{2019MNRAS.486..981G} simulations. Our profile peaks at higher values at $r =0$ and decreases to underdense values at r > 2 Mpc, while the \citet{2019MNRAS.486..981G} profile approaches a constant overdensity level of approximately 3 at $r =2-3$ Mpc.
In any case, this further demonstrates the importance of focusing the missing baryon search close to the filament spines.

As in the case of the temperatures above, we compared our results from the EAGLE simulation with the observational results from  \citet{2020A&A...637A..41T}, but this time for the radial distribution of the density of the filament gas. Again, within the central $r \approx 1$ Mpc both the simulations and the observations agree (see Fig. \ref{dprof_comparison.fig}). This lends confidence on the results and strengthens the simulation-based insight that it is beneficial to focus the missing baryon search to the central $r \approx 1$ Mpc regions around the filament axes.  For further discussion see Section \ref{disc}.

\subsection{Outside the filament core regions}
\label{Summary}
The diffuse baryon mass density profile of the high $\delta_{LD}$ filament sample is significantly above the background level in the inner $r = 1$ Mpc region.
The background density level, obtained with the randomised profiles, should be dominated by the voids due to their large volume filling fraction of gas (76\%) in EAGLE, based on a NEXUS+ analysis \citep{2019MNRAS.487.1607G}. Indeed, the background level is $\sim$10\% of the cosmic mean, consistent with the density distribution in the NEXUS+ voids (see Fig. \ref{NEXUS_BISOUS.fig}).
This indicates that on average at $r \approx$ 3 Mpc distance from a filament spine we enter the void domain.

Similarly, the background level of the temperatures, $\log{T(K)}$ $\sim$ 3.0-3.5, is consistent with the temperature distribution in the  NEXUS+ voids (see Fig. \ref{NEXUS_BISOUS.fig}).
However, the filament temperature profile indicates higher values of $\log{T(K)}$ $\approx$ 4.0-4.5 at $r \ge$ 3 Mpc (see Fig. \ref{tprof.fig}).
Analysis of the temperature distribution of the  particles in the full filament sample at these radii indicated that in addition to the  maximum at the expected void value of $\log{T(K)}$ $\approx$ 3,
$\approx$20\% of the particles have extremely high temperatures (in excess of $\log{T(K)} = 5$) which drive the medians of the temperature distributions  of individual filaments to high  values. As the full filament distribution is calculated using a log-normal fit of the individual medians, the whole distribution is thus biased towards high temperatures.
The imaging analysis indicated that this population of gas resides in extended hot regions around massive haloes, probably due to the most energetic shocks propagating to larger distances.
Their importance is negligible in the randomised profile sample due to the very small volume filling fraction of the haloes \citep{2019MNRAS.487.1607G}.
The consequent $\approx$20\% excess in the particle density at $r \approx$ 3 Mpc above the approximate void level ($\approx$10\% of the cosmic mean) is negligible compared to the scatter in the density profile. 

In summary, combined Bisous and NEXUS+ analysis indicates that at $r \approx$ 3 Mpc distance from a filament we enter the void domain, unless we are in a neighbourhood of a massive halo.

\section{Discussion}
\label{disc}

\subsection{ Limitations of EAGLE}
 We showed that hot and diffuse intergalactic gas within cosmic filaments makes up a significant fraction of the baryons in the EAGLE simulation at redshift z$ \sim$0. However, several EAGLE related limitations could potentially have significant consequences  on our results.

  The volume covered by the simulation is relatively small (100$^{3}$ Mpc$^{3}$), suppressing the possible larger structure formation modes.  Thus we were not able to study the potential variation of the cosmic baryon heating related to the longest filaments ($> 50$ Mpc) missing in EAGLE.  However, a recent filament study on large volume simulations \citep[Illustris TNG300, 300$^{3}$ Mpc$^{3}$, and Magneticum, 500$^{3}$ Mpc$^{3}$,][]{2020A&A...641A.173G} has shown that the number of filaments decreases rapidly as a function of filament length, with much less than 1\% of filaments being larger than 50 Mpc. Ignoring such a small sample will not significantly affect our results.

  On the other hand, the resolution of the simulation limits the minimum scale of structures. The mass resolution in EAGLE is enough to resolve the Jeans scales for the warm interstellar medium (of the order of kiloparsecs, \citealt{2015MNRAS.446..521S}), whereas the filamentary structure and the shocks heating the IGM are part of the large scale structure \citep[of the order of megaparsecs, see e.g][]{2003ApJ...593..599R}. Thus, we do not expect the resolution limits to affect our results. 

In addition, simulations may fail to reproduce all the relevant thermodynamic interactions that take place during structure formation. While the IGM in filaments is presumably heated mainly via accretion shocks driven by gravity \citep[e.g.][]{2003ApJ...593..599R}, the winds driven by feedback processes from star formation and AGN that are included in EAGLE are also thought to affect the WHIM \citep[e.g.][]{2012MNRAS.425.1640T}. However, these galactic winds are poorly constrained observationally. Cosmic rays, which are not modelled in EAGLE, may or may not have a significant impact on the heating process \citep{2020arXiv200402897H}. Nonetheless, while cosmic rays could potentially change the thermal structure of the CGM around the galaxies they are expelled from, the extent of these rays is limited. Indeed, cross-correlation of the extragalactic gamma-ray background with SZ maps imply that cosmic rays cannot affect for a significant fraction of the total pressure in clusters of galaxies \citep{2020PhRvD.101j3022S}. Thus, cosmic rays are not expected to alter the intergalactic gas studied here. 

Another factor that might have an effect on the results presented here is the cooling of the IGM. While the subgrid radiative cooling is implemented in EAGLE in a sophisticated, element-by-element manner, uncertainties in the intergalactic metallicities may lead to an under- or overestimate of the cooling rates. If, in reality, the metal abundances in filaments are much higher than simulated in EAGLE, the cooling rate of the IGM would be higher. This could lead to an overestimate of the missing baryon population in filaments based on EAGLE. However, the intergalactic metallicity in filaments is not well known. Adopting the common assumption of 0.1 times the Solar metallicity, filamentary gas in the hot WHIM temperature ( $ \log{T(K)} > 5.5$) and density range ($\delta \approx 1-50$, see Fig. \ref{dprof.fig}) has a cooling time longer than the Hubble time \citep{2009MNRAS.393...99W}. Assuming the extreme case of Solar metallicities in the intergalactic gas, the cooling time scale is still above or comparable to the Hubble time. Thus, any filamentary IGM that reaches these high temperatures is expected to remain hot.

 Further reassurance is given by the good agreement of our temperature and density profiles with recent observations of the tSZ effect in filaments \citep[][see Figs. \ref{tprof_comparison.fig} and \ref{dprof_comparison.fig}]{2020A&A...637A..41T}. This agreement would be highly unlikely if the EAGLE filaments were unrealistic. Thus, albeit future simulations may implement improvements to account for all the limitations mentioned above, we do not expect significant changes in our results.

\subsection{Bisous as a missing baryon finder}
We found that with our current version of the Bisous formalism, we can potentially locate a large fraction of the missing baryons by tracing them with the cosmic filaments detected using the observational spectroscopic galaxy catalogues. Based on the analysis of the EAGLE data it is beneficial to concentrate the observational missing baryon search (via e.g. the SZ effect, X-ray absorption and emission) within $\sim$ 1 Mpc distance from the axes of the cosmic filaments with the highest (observable) galaxy luminosity densities, since in these regions the diffuse baryon density exceeds ten times the cosmic mean and the baryon temperature is significantly above the approximate hot WHIM threshold of $\log{T(K)} = 5.5$ ( see Figs. \ref{tprof.fig} and \ref{dprof.fig}). On the other hand, the average Bisous filament radius (based on galaxy distributions only,  i.e. independent of the gas, see Section \ref{filres}) in our current implementation of the Bisous method is also $\sim$ 1 Mpc. Thus, there is not much room for relaxing the Bisous filament volume boundary condition (i.e. lowering the visit map value from 0.05, see Section \ref{nexus}) since it would add very little hot WHIM while increasing the amount of gas with low densities and temperatures. On the other hand, a stricter visit map criterion would result in using smaller extraction radii and consequently in less hot WHIM being covered. Thus, the current Bisous setting is close to optimal for the missing baryon search.

\subsection{Simulations versus observations}
We modelled and reported the spatial distribution of the basic thermodynamic quantities (density and temperature) of the hot diffuse gas in the filamentary environments as indicated by the EAGLE simulations.
This may serve as a starting point for producing predictions for observational signals to be used for planning future observations and interpreting the measured signals. We will carry out such work in the near future for the optimal SDSS filament sample identified in this work (see Section \ref{classification}).

 \subsubsection{The SZ effect}
Perhaps the most promising observational avenue for detecting the missing baryons is to utilise the SZ effect imposed on the Cosmic Microwave Background by the hot gas in the filaments \citep[e.g.][]{2019MNRAS.483..223T,2019A&A...624A..48D,2020A&A...637A..41T}.
The data from the Planck mission \citep{2016A&A...594A..22P} currently  has the best combination of the sky coverage and sensitivity for this work. Stacking the Planck data around the axes of the cosmic filaments for a large enough sample may enable a statistically significant signal to be measured.
Among the current spectroscopic galaxy surveys the Sloan Digital Sky Survey has the most optimal combination of sky coverage and sensitivity for producing a large data base for detecting the cosmic filaments.

The stacking of Planck tSZ and CMB-lensing data on a different sample of SDSS filaments than ours was carried out by \citet{2020A&A...637A..41T}. While \citet{2020A&A...637A..41T} used filaments constructed with the DisPerSE method and a length criterion of $l = 30-100$ Mpc, we used the Bisous method which yielded a maximum filament length of 35 Mpc. This may lead to a comparison between two distinct filament populations, as \citet{2020A&A...641A.173G} showed that in simulations longer ($l > 20$ Mpc) DisPerSE filaments are thinner and have lower galaxy densities. Thus, the Bisous high-luminosity density sample might correspond to high galaxy density, i.e. shorter and wider filaments. We keep this caveat in mind when comparing the two works below.

Within the central $r \approx 1$ Mpc, the EAGLE temperatures and baryon densities are consistent with those derived from the Planck data. This raises the question whether this is compatible with the fact that the two filament samples are chosen differently. Perhaps the central galaxy density difference between the short and long filaments \citep{2020A&A...641A.173G} is relatively small, and consequently the difference in gas properties is within the larger error bars coming from the Planck and EAGLE analysis. In that case, the consistency between EAGLE and Planck is not a co-incidence but rather lends support to the results.

However, at $r  \gtrsim 1$ Mpc there are clear differences: the EAGLE density and temperature profiles are steeper. In particular, EAGLE does not support the isothermality assumption beyond $r \approx 1$ Mpc, while the isothermal profile derived by \citet{2020A&A...637A..41T} extends up to $r = 5$ Mpc. If one was to use a radially decreasing temperature profile in the Planck analysis, the density profile should become even flatter to yield the same SZ signal.  Possibly, the difference in the samples of filaments may explain the discrepancy at larger radii. At larger radii the observed SZ signal is dominated by the PSF scatter of the signal originating from the central $r \approx 1$ Mpc in the filaments.  Also, the total signal at larger radii is background-dominated. Thus, the shape of the SZ profile may not be very accurately determined observationally. The possible bias in the temperature and density profiles at larger radii may remain unnoticed due to their very small statistical weight on the fitting procedure.

\subsubsection{X-rays}

With ATHENA it may become possible to measure a significantly large sample of X-ray absorption signals from high ions (e.g. \ion{O}{VII}) embedded in the WHIM within filaments in the spectra of blazars behind them.
Analyses of the EAGLE and IllustrisTNG simulations \citep{2019MNRAS.488.2947W,2020MNRAS.498..574W,2018MNRAS.477.450N} have shown that
the hot gas tracers \ion{O}{VII} and \ion{O}{VIII} appear densely packed around the galaxies, while the lower density phase follows the large scale structure. 
A recent work on SIMBA simulations \citep{Borrow_2019} found that AGN jets may carry 10\% of the baryons up to 4 Mpc distances from the host galaxies and thus enrich the intergalactic medium with metal ions. We will extend the above works and pave the way for ATHENA by examining and modelling the spatial distributions of the high ion densities in the EAGLE simulation within the cosmic filaments in a future work.

Once we have mapped the metals within the filaments in the EAGLE simulation, we can produce predictions for the soft X-ray emission. The currently operational eROSITA satellite may have the potential to detect the relatively faint X-ray emission form a large number of stacked 4MOST
filaments.

\section{Conclusions}
In this work we presented a novel method for localising the missing baryons around cosmic filaments and estimating their spatio-thermal properties. 
For developing our method, we chose to use the state-of-the-art hydrodynamical EAGLE simulation, as it provides information on both galaxies and baryon gas. Here we summarise our results:

\begin{itemize}
\item We identified the filamentary structure within the EAGLE simulation using the previously published NEXUS+ classification of the large scale structure environment based on the dark matter density field and by applying the Bisous model to an SDSS-like magnitude limited galaxy sample. 
\item  The diffuse hot intergalactic medium ( $\log{T(K)}$ > 5.5, outside R$_{200}$) captured by filaments amounts to $\approx 23\% - 25\%$ of the total baryon budget, or $\approx 79\%-87\%$ of all the hot WHIM. This accounts for a large fraction of the missing baryons.
\item  Filaments fill only $\approx$ 5\% of the total volume, indicating that the hot WHIM is very tightly concentrated along the filament spines. Thus our tests indicate that by focusing on the galaxy filaments detected in optical data, one can localise most of the missing baryons.
\item The hot WHIM mass in NEXUS+ and Bisous filaments differs by only $\sim$ 10\%. This indicates that the missing baryons can be reliably traced with these filament finding methods. 
\item Based on the galaxy luminosity density we selected the most optimal $\sim$10\% of the filament spines for more detailed study. The purity (the mass fraction of the missing baryons) of this high $\delta_{LD}$ sample is $\approx$ 82\%.  
\item For our high $\delta_{LD}$ sample we modelled the density and temperature profiles as a function of distance from the filament spine with analytic functions. The profiles show that in order to obtain the strongest observational signal, it is beneficial to focus on the central $\sim$ 1 Mpc from the spine.
\item Our temperature and density profiles of the filament gas are consistent with those derived from Planck tSZ and CMB lensing observations \citep{2020A&A...637A..41T} within the central $\sim 1$ Mpc regions of the filaments.
\item The filament detection methods together with the baryon density and temperature profiles derived here are applicable to observations and may thus be useful for planning future research on the missing baryons. 
\end{itemize}

\begin{acknowledgements}
  We acknowledge the support by the Estonian Research Council grants PUT246,
  IUT40-2, PRG1006, and by the European Regional Development Fund (TK133). We thank Stuart McAlpine and Till Sawala for help with the EAGLE data.
  Thanks to Davide Martizzi for his help on IllustrisTNG. We acknowledge the Virgo Consortium for making their simulation data available. The eagle simulations were performed using the DiRAC-2 facility at Durham, managed by the ICC, and the PRACE facility Curie based in France at TGCC, CEA, Bruyèresle-Châtel.
  We thank Rien van de Weygaert, Bernard J.T. Jones and Marius Cautun for their permission to use data of the NEXUS+ analysis.
  We also thank Nabila Aghanim and Hideki Tanimura for providing the density and temperature profiles derived from observations.
    \end{acknowledgements}

\bibliographystyle{aa} 

\bibliography{eaglebib} 

\end{document}